# Астрофизика одиночных нейтронных звезд: радиотихие нейтронные звезды и магнитары

С. Б. Попов[1] и М. Е. Прохоров[1]

**Аннотация**

В этой работе дается обзор современного состояния теории и наблюдений одиночных нейтронных звезд. В основном рассматриваются объекты, не проявляющие обычной пульсарной активности в радиодиапазоне. Мы отнесли к этому классу слабые рентгеновские источники, являющиеся кандидатами в одиночные нейтронные звезды, в диске Галактики и в шаровых скоплениях; компактные рентгеновские источники в остатках сверхновых; аномальные рентгеновские пульсары; источники мягких повторяющихся гамма-всплесков; объекты типа Геминги. Также мы рассматриваем родственные объекты (одиночные черные дыры и т.н. "странные" звезды) и кратко описываем современное состояние астрофизики сверхновых и связанные с этим проблемы.

Нами описываются основные процессы, связанные с эволюцией одиночных нейтронных звезд: охлаждение, затухание магнитного поля, эволюция периодов вращения, магнитосферные процессы, аккреция. Рассматривается популяционный синтез одиночных нейтронных звезд разных типов, кратко описываются "единые теории", объясняющие некоторые из перечисленных выше типов источников как последовательную эволюцию одиночных нейтронных звезд.

---

[1] *Россия, 119992 Москва, Государственный Астрономический Институт им.П.К.Штернберга*

e–mail:
  С. Б. Попов:      polar@sai.msu.ru
  М. Е. Прохоров:  mike@sai.msu.ru

# Содержание

# 1 Введение

Наблюдения нейтронных звезд (НЗ) имеют уже почти 35-летнюю историю (Хьюиш, Белл и др. 1968). Основная доля известных НЗ — радиопульсары (см. данные о недавних наблюдениях в Лоример и др. 2000, Д'Амико и др. 1999 и каталог в Тейлор и др. 1993 и на сайте http://wwwatnf.atnf.csiro.au/research/pulsar/). В течение первых лет изучения НЗ установилась картина эволюции этих объектов, в которой роль своеобразного стандарта выполнял пульсар в Крабовидной туманности. Это НЗ, имеющая сейчас период вращения около 0.033 секунд и магнитное поле $\sim 5.2 \cdot 10^{12}$ Гс. Период увеличивается с характерным значением $\dot p \sim 4.16 \cdot 10^{-13}$ с/с в грубом соответствии с магнито-дипольной формулой. Начальный период пульсара был порядка 0.020 секунд. Возраст НЗ — около 1000 лет.

Новые данные наблюдений, в первую очередь полученные в рентгеновском диапазоне, существенно дополняют эту картину. Уже в середине 90-х годов (см. Каравео и др. 1996) начало формироваться мнение о радиотихих НЗ как о "новой астрономической реальности". Теперь пульсар в Крабовидной туманности потерял свою роль полномочного представителя всей популяции НЗ, устоялся термин *радиотихие* одиночные НЗ. Об этом и пойдет речь в данной статье.

Наблюдения в рентгеновском диапазоне, в первую очередь на спутнике РОСАТ (см. например Мотч 2000), показали наличие популяции источников (а возможно и нескольких различных популяций, см. ниже), отождествляемых с одиночными НЗ. Есть основания предполагать, что часть из этих объектов генетически связана с т.н. аномальными рентгеновскими пульсарами и источниками повторяющихся гамма-всплесков. Все это делает одиночные НЗ исключительно интересным объектом исследования.

Здесь мы исключаем из рассмотрения объекты, являющиеся обычными радиопульсарами (см. общий обзор в статье Бескина 1999, а обзор рентгеновских наблюдений радиопульсаров см. в работе Бекера 2000).

Мы обсудим шесть типов источников, в которых с большей или меньшей степенью достоверности находятся одиночные НЗ: слабые рентгеновские источники в диске Галактики, аномальные рентгеновские пульсары (АРП), источники мягких повторяющихся гамма-всплесков (МПГ), компактные рентгеновские источники в остатках сверхновых, слабые рентгеновские источники в шаровых скоплениях, объекты типа Геминги. Разумеется, некоторые источники могут входить сразу в две популяции (например, АРП в остатке сверхновой).

Мы описываем физические процессы, непосредственно связанные с эволюцией и природой одиночных НЗ: эволюция периодов НЗ, аккреция на одиночные НЗ, затухание магнитного поля НЗ, охлаждение НЗ, атмосферы НЗ и др. Также в данном обзоре мы кратко обсуждаем различные гипотезы, объясняющие происхождение и свойства одиночных НЗ, не являющихся классическими радиопульсарами.

Недавние обзоры теоретических моделей различных физических процессов, связанных с НЗ (особенно с внутренним строением НЗ), читатель может найти в материалах многочисленных недавних конференций, например, Тренто-2000 ("Physics of neutron star interiors"), Бонн-1999 (Proceedings of IAU Coll. 177), Токио-1997 ("Neutron stars and pulsars"), Эльба-1999 ("The relationship between neutron stars and supernova remnants"), книгах – Шапиро и Тьюколски 1985, Липунов 1987, Майкель 1991, Саакян 1995, а также в обзорах (Яковлев и др. 1999, Ланглуа 2000, Клужняк 2000, Яковлев, Каминкер и др. 2001, Картер 2001).

Мы приводим обширную библиографию по всем упомянутым в тексте проблемам.

Обзор также доступен на сайте **http://xray.sai.msu.ru/~polar/ns_review/**.

## 2 Актуальные проблемы астрофизики одиночных нейтронных звезд

Основной задачей изучения одиночных НЗ является *определение распределений начальных параметров и законов эволюции этих объектов*. Дополнительной задачей является выяснение причин возникновения соответствующих распределений.

Сегодня можно сформулировать следующие актуальные проблемы и направления исследований, существующие в этой области:

**I. Наблюдения НЗ**

- Наблюдения в рентгеновском диапазоне: наблюдения одиночных НЗ всех типов, поиск спектральных особенностей, исследования остатков сверхновых (поиск компактных объектов);

- Наблюдения в радио диапазоне: поиск радиоизлучения от одиночных НЗ всех типов;

- Наблюдения в ИК диапазоне: поиск слабых компаньонов и/или остаточных аккреционных дисков (аномальные рентгеновские пульсары, источники мягких повторяющихся гамма всплесков);

- Наблюдения в оптическом диапазоне: регистрация теплового излучения НЗ и поиск слабых компаньонов и/или остаточных аккреционных дисков;

**II. Физика НЗ**

- Расчеты тепловой эволюции НЗ;

- Расчеты спектров НЗ с учетом атмосфер различного состава;

- Процессы в магнитосферах: электродинамика магнитосфер, перенос излучения;

- Расчеты затухания магнитного поля НЗ;

- Аккреция на одиночные НЗ из межзвездной среды и из околозвездных остаточных дисков: темп аккреции, перенос момента и т.д.;

- Расчеты эволюции периодов вращения НЗ;

- Популяционный синтез НЗ, попытки построения "единых теорий";

**III. Другие близкие области исследований**

- Механизмы взрывов сверхновых и возвратная аккреция (fall-back); определение начальных параметров НЗ;

- Одиночные черные дыры: аккреция, возможные наблюдательные проявления;

- Странные звезды;

Среди наблюдательных задач (практически во всех диапазонах) можно особо выделить исследования переменности (особенно определение и мониторинг производной периода вращения), независимое определение магнитного поля и возраста (без использования данных о замедлении), исследования ассоциации НЗ с остатками сверхновых и поиск новых кандидатов. Большой интерес представляют также наблюдения в самых жестких диапазонах, т.к. относительно близкие радиопульсары, чей пучок излучения не попадает на Землю, могут наблюдаться как гамма-источники (Реймер и др. 2001, Гренье 2000, Гренье и Перро 2001). Отметим также, что для многих из известных источников нет точных определений расстояний и, разумеется, скоростей (об определении параллаксов радиопульсаров см., например, Тоскано и др. 1999, Чаттерье и др. 2001, Брискен и др. 2002 и ссылки там).

Ниже мы рассматриваем эти задачи подробнее.

Таблица 1: Слабые рентгеновские источники в диске Галактики (из работы Тревес и др. 2000 с дополнениями)

| Название источника | Поток[a] [отсч./с] | Эфф. темп. [эВ] | $N_H$ [$10^{20}$ см$^2$] | $\log f_X/f_V$ [b] | Период [с] |
|---|---|---|---|---|---|
| RX J185635-3754 | 3.64 | 57 | 2 | 4.9 | — |
| RX J0720.4-3125 | 1.69 | 79 | 1.3 | 5.3 | 8.37 |
| RBS1223 (1RXS J130848.6+212708) | 0.29 | 118 | $\sim 1$ | $> 4.1$ | 5.15 |
| RBS1556 (RX J1605.3+3249) | 0.88 | 100 | $< 1$ | $> 3.5$ | — |
| RX J0806.4-4123 | 0.38 | 78 | 2.5 | $> 3.4$ | 11.37 |
| RX J0420.0-5022 | 0.11 | 57 | 1.7 | $> 3.3$ | 22.7 |
| RBS1774 (1RXS J214303.7+065419) | 0.18 | 90 | 4.6 | $> 3$ | — |

[a] Отсчеты для спутника РОСАТ.
[b] $f_X$ и $f_V$ — потоки в рентгеновском (РОСАТ) и оптическом диапазонах, соответственно.

### 2.1 Наблюдения НЗ

Наиболее обширные и длительные наблюдения НЗ — наблюдения радиопульсаров в радиодиапазоне — не входят в данный обзор. Из прочих данных самыми подробными являются рентгеновские и гамма наблюдения. В радио-, ИК- и оптическом диапазонах о НЗ получено заметно меньше сведений.

#### 2.1.1 Наблюдения в рентгеновском и гамма диапазонах

Для одиночных радиотихих НЗ наиболее важным оказался рентгеновский диапазон. В этом пункте мы рассмотрим близкие радиотихие НЗ, слабые рентгеновские источники в шаровых скоплениях, источники в остатках сверхновых, аномальные рентгеновские пульсары и источники мягких повторяющихся гамма-всплесков.

Спутником РОСАТ открыто уже семь радиотихих нейтронных звезд. Еще один кандидат (MS 0317) был открыт ранее на обсерватории "Эйнштейн", однако в последнее время стало очевидным, что он является внегалактическим объектом (мы благодарим Г.Г. Павлова за соответствующее замечание). Это относительно яркие объекты ($> 0.1$ отсчета в секунду). Два источника зарегистрированы в оптическом диапазоне (RX J1856 и RX J0720). Для остальных существуют только верхние пределы (см. таблицу 1). По отношению рентгеновского и оптического потоков можно считать доказанным, что эти источники являются одиночными нейтронными звездами.

В начале 90-х годов ожидалось, что спутник РОСАТ увидит большое количество одиночных старых аккрецирующих НЗ (Тревес и Колпи 1991, Блаез и Раджагопал 1991, Блаез и Мадау 1993, Колпи и др. 1993). Последующие наблюдения на этом спутнике показали наличие лишь небольшой популяции слабых рентгеновских источников в диске Галактики, которые по-видимому являются одиночными НЗ (охлаждающимися или аккрецирующими) (см. Хаберл и др. 1998, Тревес и др. 2000, Нойхойзер, Трюмпер 1999; последний кандидат описан в работе Зампьери и др. 2001). Основным аргументом в пользу такой интерпретации является аномально высокое отношение рентгеновской светимости к оптической ($\log f_X/f_V > 3$). Отсутствие радиоизлучения, а также тепловой спектр свидетельствуют о том, что объекты не являются классическими радиопульсарами.

У четырех объектов "великолепной семерки" наблюдаются пульсации рентгеновского потока с периодами 5-20 секунд. Длительные (450 000 сек) наблюдения наиболее изученного источника RX J185635-3754 (Рэнсом и др. 2001) показали отсутствие пульсаций вплоть до нескольких процентов.

Для двух источников измерены производные периода. В работе Дзане и др. (2002) получено значение $\dot p \sim 10^{-14}$ для источника RX J0720.4-3125. Хамбарян и др. (2001) приводят значение $(0.7 - 2.0) \cdot 10^{-11}$ для источника RBS1223.

Пфал и Раппопорт (2001) предположили, что некоторые из слабых источников в шаровых скоплениях могут быть старыми аккрецирующими одиночными НЗ. В настоящее время спутником Чандра открыто множество подобных источников и их количество стремительно продолжает расти (см. также Вербунт 2001). Для них характерны светимости порядка $10^{31} - 10^{34}$ эрг/с. Эти объекты концентрируются к центрам скоплений. Источники характеризуются относительно мягкими спектрами $kT \sim 0.1$–$0.5$ кэВ.

Популяционный синтез (см. ниже) показал, что гипотеза Пфала и Раппапорта не противоречит стандартной картине эволюции НЗ. Таким образом необходимы дальнейшие наблюдения (в первую очередь исследования переменности на разных масштабах) для определения природы этих объектов.

Для понимания ключевого вопроса о начальных параметрах НЗ крайне важным классом объектов являются источники в остатках сверхновых (обзор по механизмам вспышек см. в Янка и др. 2001). В последнее время открыто несколько компактных источников в остатках сверхновых (см. Павлов и др. 2001а). По всей видимости это молодые НЗ, многие из которых не являются радиопульсарами (см. таблицу 2). Количество радиотихих НЗ в остатках сверхновых уже сравнимо с количеством радиопульсаров в остатках. По этой теме существует обширная библиография (см. таблицу 2).

Можно выделить три основные гипотезы, объясняющие появление компактных рентгеновских источников в остатках сверхновых: тепловое излучение поверхности молодой горячей НЗ, нетепловое излучение молодого пульсара (в этом случае следует ожидать если не регистрации самого пульсара, то хотя бы возникновения плериона), возвратная аккреция на молодую НЗ (или черную дыру) вещества остатка сверхновой (fall-back).

Важными наблюдательными фактами для интерпретации природы источников являются наличие периодов и переменность рентгеновского потока. Периоды АРП и МПГ лежат в диапазоне 5-12 секунд. Другие источники обладают короткими периодами (например p=0.325 с и производная периода $\dot p \sim 7.1 \cdot 10^{-12}$ у источника в Kes 75 (Мерегетти и др. 2002а), p=0.424 с у источника в G296.5+10, для этого объекта также измерена производная периода $\dot p \sim (0.7-3) \cdot 10^{-14}$ (Павлов и др. 2002)). Источники в RCW 103 и G29.6+0.1 показывают существенную переменность рентгеновского излучения на больших временах, поток при этом может изменяться на порядок.

Рассмотрим трудности в интерпретации радиотихих источников на конкретном примере *Cas A*.

Остаток сверхновой Cas A является хорошо известным объектом, но природа компактного объекта в настоящее время неясна.

Расстояние до остатка порядка 3.4 кпк. Его радиус около 2 пк. Возраст остатка оценивается примерно в триста лет (есть указания, что вспышка сверхновой наблюдалась Флемстедом в 1680 г.). Это самый молодой из известных остатков сверхновых в нашей Галактике.

Наблюдения на спутнике Чандра дали много новых данных. Собственно компактный источник был открыт во время первых наблюдений на этом спутнике в августе 1999 г. (здесь мы следуем работам Павлова и др. 2000, Чакрабарти и др. 2001).

Компактный источник расположен практически в центре остатка. Оптические и ИК наблюдения дают только верхние пределы на излучение компактного объекта.

Сложности начинаются уже при определении светимости компактного источника. В зависимости от модели спектра (чернотельный или степенной) светимость оценивается от $\sim 2 \cdot 10^{33}$ эрг с$^{-1}$ до $\sim 6 \cdot 10^{35}$ эрг с$^{-1}$. Важной особенностью является маленькая площадь ($\sim 1$ км$^2$) излучающей поверхности во всех вариантах интерпретации спектра. Эти характеристики не дают возможности однозначно определить природу компактного источника и механизм излучения.

Радиопульсары в остатках сверхновых являются подклассом наиболее молодых пульсаров, и потому чрезвычайно важны для определения начальных параметров НЗ. Однако, до сих пор не ясно, какая доля сверхновых порождает радиопульсары. Вопросы ассоциации радиопульсаров с остатками сверхновых с наблюдательной точки зрения детально рассматривались в обзорах Каспи (1996), Лоримера и др. (1998), Готтхелфа и Васиштта (2000) и статьях Брауна и др (1989), Камило и др. (2001), Нараяна и Шаудта (1988). Теоретические

Таблица 2: Компактные рентгеновские источники в остатках сверхновых (из работы Чакрабарти и др. 2000 с дополнениями из Хелфанд 1998)

| Название источника и остатка | Тип остатка | Период, с | Ссылки |
|---|---|---|---|
| Cas A | Оболоч. | — | Чакрабарти и др. (2000) |
| 1E 121348-5055 (RCW 103) | Оболоч. | — | Каспи и др. (1998) |
| 1E 0820-4247 (Pup A) | Оболоч. | — | Петре и др. (1996) |
| 1E 1207.4-5209 (G296.5+10) | Оболоч. | 0.424 | Мерегетти и др. (1996) |
|  |  |  | Павлов и др. (2000) |
| AX J0851.9-4617.4 (G266.2-1.2) | Оболоч. | — | Слейн и др. (2001) |
|  |  |  | Мерегетти (2000) |
| PSR J1846-0258 (Kes 75) | Композ. | 0.325 | Готтхелф и др. (2000) |
| 1E 1438.7-6215 (RCW 86) | Оболоч. | — | Винк и др. (2000) |
| RX J0002+6246 (G117.7+0.6) | Оболоч. | 0.242 | Бразиер, Джонстон (1999) |
| RX J2020.2+4026 (G78.2+2.1) | Оболоч. | — | Бразиер, Джонстон (1999) |
| RX J0201.8+6435 (3C58) | Плерион | 0.065 | Мюррей и др. (2002) |
| RX J0007.0+7302 (CTA1) | Плерион | — | Чакрабарти и др. (2000) |
| Название источника и остатка | Тип источника | Период, с | Ссылки |
| 1E 1841-045 (Kes 73) | АРП | 11.8 | Мерегетти (1999) |
| AX J1845-0258 (G29.6+0.1) | АРП | 6.97 | Генслер и др. (1999) |
|  |  |  | Мерегетти (1999) |
| 1E 2259+586 (CTB 109) | АРП | 6.98 | Мерегетти (1999) |
| SGR 1900+14 (G42.8+0.6) | МПГ | 5.16 | Харлей (1999) |
| SGR 0526-66 (N49) | МПГ | 8 | Харлей (1999) |
| SGR 1627-41 (G337.0-0.1) | МПГ | 6.41 | Харлей (1999) |

расчеты числа таких ассоциаций можно найти в работах Генслера и Джонстона (1995а,б,в). Трудности, связанные с определением возрастов пульсаров, недавно обсуждались на примере B1757-24 (Генслер, Фрейл 2000). АРП и МПГ в остатках сверхновых изучались Марсденом с соавторами (1999). Ранний обзор по компактным рентгеновским источникам в остатках сверхновых можно найти в работе Бразиер и Джонстона (1999). По всем этим вопросам было проведено несколько крупных международных конференций (см., например, "The relationship between neutron stars and supernova remnants"). На русском языке детальный обзор остатков сверхновых приведен в книге Лозинской (1986), а также в обзоре Блинников и др. (1987).

Пожалуй самыми "модными" объектами из всех, рассматривающихся в нашем обзоре являются АРП (Мерегетти, Стелла 1995, ван Парадайз и др. 1995) и МПГ (Мазец и др. 1979, см. также обзор ранних наблюдений в Мазец, Голенецкий 1987).

МПГ представляют собой объекты, демонстрирующие случайные (непредсказуемые) периоды вспышечной активности в мягком ($< 100$ кэВ) гамма-диапазоне (см. каталог МПГ в статье Аптекарь и др. 2001). Периоды активности длятся от дней до месяцев. Вспышки соответствуют светимости выше эддингтоновской для объекта солнечной массы. Иногда происходят гигантские вспышки, которые характеризуются более жестким спектром и энергией в тысячи раз больше, чем выделяется в обычной вспышке.

В спокойном состоянии МПГ наблюдаются как относительно мягкие рентгеновские источники. Всего на данный момент известно 4 таких источника: SGR 1900+14 (p=5.16 с), SGR 1806-20 (p=7.47 с), SGR 1627-41 (p=6.4 с), SGR 0525-66 (p=8 с). (см. обзоры в Израел и др. 2001, Мерегетти 1999, Харлей 1999, Томпсон 2000). Некоторые из них находятся в остатках сверхновых, однако степень достоверности генетической связи между МПГ и остатками остается под вопросом (см. таблицу 2). Источник SGR 1806-20 находится в скоплении массивных звезд (Эйкенберри 2002).

АРП были выделены в отдельный класс в 1995 году. Они характеризуются близкими периодами порядка 6-12 секунд (1E 1048.1-5937 — 6.44 с; 4U 0142+61 — 8.69 с; 1E 1841-045 — 11.77 с; 1E 2259+586 — 6.98 с; 1RXS J170849.0-400910 — 10.99 с; AX J1845-0258 — 6.97 с),

низкими светимостями ($\sim 10^{35}$ эрг с$^{-1}$) и более мягким спектром, чем у обычных рентгеновских пульсаров в тесных двойных системах, стабильной светимостью на больших масштабах времени (обычно для рентгеновских пульсаров в тесных двойных системах характерна заметная переменность), постоянным замедлением (т.е. отсутствием эпизодов уменьшения периода вращения) и отсутствием данных о наличии второго компонента системы (см. обзоры Мерегетти 1999, Томпсона 2000 и Мерегетти и др. 2002б).

Сейчас уже не вызывает больших сомнений то, что АРП являются одиночными НЗ (не исключены только очень маломассивные компаньоны). Однако, природа светимости (аккреция или тепловое излучение горячей звезды) остается неизвестной. Вероятно наиболее важную информацию можно получать из детальных данных об изменении периодов этих объектов (Гавриил, Каспи 2002, Пол и др. 2000), а также из спектральных характеристик (особенно во время вспышек см. Марсден и Вайт 2001, Перна и др. 2001). Для получения этой информации необходим мониторинг (см. Готтхелф и др. 2001, Гавриил и др. 2001).

В последнее время появились результаты по рентгеновской спектроскопии АРП и МПГ. Спутник Чандра наблюдал АРП 4U 0142+61 (Джюэт и др. 2002). Не было найдено никаких указаний на присутствие в спектре эмиссионных или абсорбционных линий. Если источник является магнитаром, то эти данные накладывают существенные ограничения на модель атмосферы.

Особое место среди одиночных радиотихих НЗ занимает Геминга (однако, на очень низких частотах удалось зарегистрировать и радиоизлучение от этого объекта, см. Кузьмин и Лосовский 1997, Малофеев и Малов 1997, Шитов и Пугачев 1998). Объект Геминга (см. Каравео 2000, Биньями, Каравео 1996) был открыт в гамма диапазоне в 1973 г. на спутнике SAS-2. В 1992 г. Хальперн и Холт объявили об открытии периода 237 мс. Источники данного типа очень трудно обнаружить, поэтому доля подобных НЗ неизвестна даже примерно. Вероятно, что Геминга это радиопульсар, чей основной пучок излучения не попадает на Землю, поэтому удается наблюдать лишь низкочастотное радиоизлучение. Ожидается, что будущие спутники (GLAST и др.) смогут увидеть в жестком диапазоне сотни НЗ такого типа в Галактике. Сегодня известен обдин объект очень похожий по всоим свойствам не геминге — 3EG J1835+5918 (Мирабал, Гальперн 2001)

### 2.1.2 Наблюдения в радио диапазоне

В настоящее время основная масса нейтронных звезд наблюдаются как радиопульсары. Сейчас число известных источников этого типа перевалило за тысячу, а в будущем с созданием километровой антенной решетки станет возможным наблюдение практически всех радиопульсаров в Галактике, чей пучок направлен на нас (Браун 1996). В связи с этим оказывается очень важным пытаться пронаблюдать все НЗ в этом диапазоне.

В большинстве моделей радиоизлучения НЗ конус излучения шире в низкочастотной области, поэтому активные поиски радио сигналов от "нестандартных" НЗ ведутся в ПРАО АКЦ ФИ РАН на низких частотах (порядка 110 Мгц). Уже зарегистрировано радиоизлучение Геминги (Кузьмин и Лосовский 1997, Малофеев и Малов 1997, Шитов и Пугачев 1998), одного источника повторяющихся гамма-всплесков (Шитов и др. 2000) и одного аномального рентгеновского пульсара (Малофеев, Малов, см. astro-ph/0106435 стр.31). В связи с уникальностью радиотелескопа ПРАО эти результаты пока не имеют надежного независимого подтверждения, кроме регистрации Геминги (см. Ватс и др. 1997). Отсутствие излучения на более высокочастотной части радио диапазона может свидетельствовать о крутом спектре.

### 2.1.3 Наблюдения в ИК диапазоне

Наблюдения в ИК диапазоне относятся в основном к аномальным рентгеновским пульсарам. Для объяснения этих источников предложено существование остаточных аккреционных дисков (см. ниже раздел, посвященный аккреции). Для выбора между существующими моделями и необходимы ИК наблюдения.

Для источника AXP 1E2259+58.6 получены данные, указывающие на наличие ИК объекта (Халлеман и др. 2001). Клозе и др. (2002) сообщают об отрицательных результатах ИК наблюдений источника SGR 1900+14.

Пока полученные результаты свидетельствуют об отсутствии аккреционных дисков или маломассивных компонентов у АРП и МПГ. Однако, необходимы дальнейшие исследования.

### 2.1.4 Наблюдения в оптическом диапазоне

Наблюдения одиночных НЗ всех типов в оптическом диапазоне представляют большой интерес (см. Каравео 2000).

Начнем с семи одиночных радиотихих НЗ, открытых спутником РОСАТ. Эти объекты были выделены как возможные НЗ именно по большому отношению рентгеновской светимости к оптической (см. таблицу 1). Лишь два источника достоверно зарегистрированы в оптике. Для остальных известны только верхние пределы.

Возможной интерпретацией рентгеновского потока от объектов "великолепной семерки" является тепловое излучение поверхности молодой НЗ. В связи с этим важно зарегистрировать тепловое излучение в других спектральных диапазонах.

Наблюдения НЗ активно ведутся на Космическом телескопе. Так по наблюдениям SGR 0526-66 (Каплан 2001а) получены важные ограничения на обе модели (аккреционную и магнитарную).

Много наблюдательных данных получено по источникам RX J1856 и RXJ0720. В частности, для RX J1856 удалось определить параллакс (первым, но с существенной ошибкой, параллакс получил Волтер (2001), уточненнение этого значения было сделано в (Каплан и др. 2001б), а затем в (Волтер, Латтимер 2002)). Расстояние до объекта равно $117 \pm 12$ пк.

Предложено оптическое отождествление для аномального пульсара 4U 0142+614 (Халлеман и др. 2000). Для нескольких источников получены важные верхние пределы (Израел и др. 2001).

Особое место занимают оптические наблюдения Геминги (см. Биньями и Каравео 1996).

Кроме наблюдения оптического излучения от самой НЗ важным является наблюдение ее самых близких окрестностей. Так наблюдение туманностей вокруг радиопульсаров позволило сделать заключение о скорости их движения и о том, что на измеренном расстоянии от них истекают потоки частиц, а не электромагнитного излучения. Ван Керквийк и Кулкарни (2001) обнаружили в $H_\alpha$ кометообразную туманность вокруг источника RX J1856. Это открытие является существенным аргументом против аккреционной интерпретации излучения данного объекта.

В нашей стране оптическими наблюдениями НЗ активно занимаются две группы на 6-метровом телескопе в САО РАН (Курт и др. 1997, Копцевич и др. 2001, Голден и др. 2000 и Бескин, Неустроев 2001). Пока положительный результат получен только для радиопульсаров и Геминги, но наблюдения продолжаются. Оригинальными результатами последней группы было обнаружение отсутствия микроструктуры оптического импульса и наличие оптического излучения (для пульсара в Крабе) в межимпульсном промежутке.

## 2.2 Физика НЗ

### 2.2.1 Расчеты тепловой эволюции НЗ

Нейтронные звезды рождаются на заключительных стадиях эволюции массивных ($M > 8$–$10 M_\odot$) звезд в результате потери устойчивости и коллапса их ядер. Этот процесс сопровождается вспышкой сверхновой. Нейтронные звезды рождаются очень горячими, до 90% выделяющейся при коллапсе ядра гравитационной энергии (порядка $10^{53}$ эрг) выделяется в виде очень мощного потока нейтрино, испускаемого горячей молодой нейтронной звездой, в первые несколько секунд или, самое большее, в первые 10–20 секунд (Пракаш и др. 2001). Для близких сверхновых (Галактика или галактики местной группы) нейтринная вспышка с помощью современной аппаратуры может быть зафиксирована непосредственно. Единственным пока случаем прямой регистрации нейтрино от вспышки сверхновой была SN1987a в Большом Магеллановом Облаке (см. Хирата и др. 1987, Бионта и др. 1987, Алексеев и др.

1988 и обзор Имшенник, Надежин 1988). Хотя НЗ испускает нейтрино и на более поздних этапах своего остывания, но их поток на много порядков меньше начальной вспышки и не может быть зарегистрирован современными средствами.

Далее в НЗ действуют два различных механизма охлаждения: нейтринный — за счет излучения нейтрино и антинейтрино из всего объема НЗ (в основном из центральной его части) и фотонное охлаждение за счет электромагнитного излучения с поверхности НЗ. Нейтринный механизм более эффективен, пока внутренняя (центральная) температура НЗ превосходит $\sim 10^8$ К, что в типичных случаях соответствует температуре поверхности НЗ $T_s > 10^6$ К. Обычно эта стадия длится $10^5$–$10^6$ лет. Современный спутниковый рентгеновский эксперимент позволяет обнаруживать только достаточно близкие и яркие, то есть молодые и горячие НЗ, поэтому ниже основное внимание будет уделено рассмотрению процессов на стадии нейтринного охлаждения.

Пионерской работой по остыванию НЗ, написанной еще *до* открытия радиопульсаров, является работа (Цурюта, Камерон 1965). Затем эта проблема исследовалась рядом авторов, но в настоящее время нет единой точки зрения по этому вопросу.

С математической точки зрения моделирование остывания НЗ сводится к решению уравнения диффузии тепла внутри звезды (Торн 1977, Левенфиш и др. 1999) с учетом объемных (нейтринное излучение) и поверхностных (фотонное охлаждение) стоков энергии. В большинстве случаев достаточно рассматривать одномерную сферически-симметричную задачу. Составными частями теории остывания НЗ являются: теплоемкость и теплопроводность ядра НЗ; величина нейтринных потерь энергии; теплопроводность коры НЗ, которая определяет связь центральной и поверхностной температур.

В первые 100–1000 лет с момента образования НЗ процессы переноса тепла внутри НЗ достаточно сложны, так как на этой стадии температуры различных внутренних частей НЗ существенно отличаются друг от друга (оболочка горячее охлаждающегося из-за нейтринного излучения ядра, см., например, Гнедин и др. 2001). Однако на этой стадии электромагнитное излучение НЗ скорее всего невозможно наблюдать из-за большой оптической толщины сброшенной во время взрыва сверхновой оболочки, которая становится прозрачной для мягкого рентгеновского излучения поверхности НЗ лишь примерно через 100 лет после взрыва. После первичной стадии тепловой релаксации ядро НЗ становится практически изотермическим, а весь перепад температуры ядра и поверхности НЗ определяется теплопроводящими свойствами коры НЗ.

Нейтрино, испускаемые горячей НЗ, образуются в ходе различных микроскопических процессов. Рассмотрим их:

- **Прямые урка-процессы** [Гамов, Шенберг (1941); Пинаев (1963)].
  Это обычные процессы бета-распада и бета-захвата, проходящие в веществе ядра НЗ.

  $$n \to p + e + \widetilde{\nu}_e, \qquad p + e \to n + \nu_e, \qquad n + e^+ \to p + \widetilde{\nu}_e.$$

  Реакции захвата и распада идут с одинаковым темпом и, таким образом, состав вещества не изменяется, а рождающиеся в каждом из процессов нейтрино и антинейтрино свободно покидают НЗ.

  Прямые урка-процессы — самый мощный источник нейтрино, но они возможны не всегда. Для того, что чтобы бета-реакции могли идти, должно выполняться так называемое "условие треугольника" $p_\mathrm{F}(n) \leq p_\mathrm{F}(e) + p_\mathrm{F}(p)$, вытекающее из закона сохранения импульса частиц. В идеальном газе вырожденных нуклонов и электронов это условие не выполняется никогда. Однако, в более реалистичных уравнениях состояния при высоких плотностях (около $2 \cdot 10^{15}$ г/см$^3$) данные реакции становятся возможными (Латтимер и др. 1991, Яковлев и др. 1999).

- **Модифицированные урка-процессы** [Чиу, Салпитер (1964)].
  Они отличаются от прямых тем, что в реакции участвует дополнительный нуклон:

  $$n + n \to n + p + e + \widetilde{\nu}_e, \qquad n + p + e \to n + n + \nu_e,$$

$$p + n \to p + p + e + \widetilde{\nu}_e, \qquad p + p + e \to p + n + \nu_e.$$

Участие в реакции дополнительной частицы снимает ограничения, налагаемые законом сохранения импульса. Данные реакции возможны в ядрах НЗ с практически любыми уравнениями состояния. С присутствием дополнительного нуклона связано то, что производимый модифицированными урка-процессами поток нейтрино заметно слабее, чем в прямых реакциях, и быстрее убывает при понижении температуры.

Именно модифицированные урка-процессы считаются основным механизмом генерации нейтрино при *стандартном* остывании НЗ.

- **Тормозное $\nu$-излучение.**
Пары нейтрино–антинейтрино могут рождаться при столкновении нуклонов в одной из следующих реакций:

$$n + n \to n + n + \nu + \widetilde{\nu}, \quad n + p \to n + p + \nu + \widetilde{\nu}, \quad p + p \to p + p + \nu + \widetilde{\nu}.$$

Зависимость выхода нейтрино от температуры в данных реакциях такая же, как в модифицированных урка-процессах, но поток для нормального (несверхтекучего) вещества на 1–2 порядка ниже (Фримен, Максвелл 1979). Однако, эти процессы могут стать важны при наличии сверхтекучести.

Рождение $\nu\widetilde{\nu}$-пар возможно, также, при рассеянии электронов (Каминкер, Хенсел 1999)

$$e + e \to e + e + \nu + \widetilde{\nu}.$$

В несверхтекучей среде этот процесс слабее всех других. Однако, он совершенно не зависит от сверхтекучести и может становиться важным после ее наступления.

- **Испускание $\nu$ при куперовском спаривании** [Флауэрс и др. 1976].
Этот процесс представляет собой испускание пары нейтрино-антинейтрино любого типа при переходе нуклона через щель в энергетическом спектре сверхтекучего вещества.

$$N \to N + \nu + \widetilde{\nu}.$$

В отсутствие сверхтекучести подобное испускание $\nu\widetilde{\nu}$-пары свободным нуклоном запрещено законами сохранения. Этот процесс был впервые предложен в 1976 в работе (Флауэрс и др. 1976), затем вновь был исследован Воскресенским и Сенаторовым (1987), но в расчетах остывания НЗ стал учитываться только с 1997 года. В общем случае данный процесс уступает по эффективности генерации нейтрино прямым урка-процессам. Однако, его учет при моделировании остывания НЗ обязателен, так как он действует в тех случаях, когда урка-процессы подавляются сверхтекучестью.

Какие из перечисленных четырех процессов действуют в конкретной НЗ с заданной массой и температурой зависит от свойств нейтронного вещества при плотностях, достигающихся в центре НЗ. Так диапазон масс НЗ и плотностей в их центрах зависит от жесткости уравнения состояния (Латтимер, Пракаш 2001). Наибольшие массы достигаются при жестких уравнениях состояния. За возможность включения прямых урка-процессов во внутренних ядрах НЗ отвечает другой параметр уравнения состояния — так называемая "асимметрия". (В центрах НЗ с не очень высокой массой и низкими центральными плотностями прямые урка-процессы всегда запрещены, но для более массивных НЗ возможность протекания данных реакций почти не зависит от жесткости уравнения состояния, см., например, (Каминкер и др. 2002).)

Присутствие протонной и/или нейтронной сверхтекучести и поведение их критических температур являются дополнительными параметрами задачи остывания горячей НЗ. Наличие и свойства сверхтекучести сильно зависят как от самого уравнения состояния, так и от метода учета многочастичных эффектов. На сегодняшний день предложенные различными авторами модели сверхтекучести перекрывают весь интересный для астрофизики диапазон свойств (см., например, рис.3 из Яковлев и др. 1999). Важность сверхтекучести для процессов остывания заключается в том, что ее наличие может частично или полностью подавлять урка-процессы и, таким образом, существенно изменять кривые остывания НЗ.

Еще одной возможностью, изученной на сегодняшний момент менее других, является появление в центре НЗ при плотности в несколько раз выше ядерной экзотических частиц. В литературе в настоящее время обсуждается несколько гипотез: рождение в центре звезды Σ- и Λ-гиперонов; образование в центре НЗ пионного или каонного конденсата (это две разные гипотезы) (об остывании НЗ с учетом пионного конденсата см. Воскресенский, Сенаторов 1986); фазовый переход к странной материи — плазме почти свободных u, d и s кварков (о странных звездах см. ниже). Любой из указанных вариантов может усилить нейтринную светимость НЗ на несколько порядков (Блашке и др. 2001).

Качественно различающиеся кривые остывания НЗ приведены на рис. 1. "Стандартной" называется кривая, определяемая модифицированными урка-процессами без сверхтекучести и прямых урка-реакций (см. параграф 2.2.1). Включение прямых урка-процессов (по прежнему в отсутствие сверхтекучести) приводит к резкому падению температуры НЗ, когда ее возраст достигает нескольких сотен лет. Это так называемое ускоренное остывание НЗ (см. параграф 2.2.1). Наличие сверхтекучести протонов и/или нейтронов вызывает сильное подавление урка-процессов (как прямых, так и модифицированных) при температурах ниже критической. Протонная сверхтекучесть обычно наступает раньше, а сверхтекучесть нейтронов сильнее всего сказывается в конце стадии нейтринного остывания НЗ. На смену подавленным механизмам излучения нейтрино через урка-процессы приходит излучение $\nu\widetilde{\nu}$-пар в сверхтекучем веществе при куперовском спаривании нуклонов (см. параграф 2.2.1). Наблюдательные данные (см. табл 3) позволяют жестко ограничить критическую температуру наступления нейтронной сверхтекучести (отметим, что поток наблюдательных данных постоянно растет в первую очередь благодаря спектральным исследованиям на рентгеновских спутниках, см. последний результат в Маршалл, Шульц 2002).

Таким образом, можно сказать, что медленнее всего спадают кривые остывания НЗ, получаемые при подавленных сверхтекучестью урка-процессах без учета куперовского механизма генерации нейтрино. Быстрее всего — кривые ускоренного остывания с прямыми урка-процессами и без сверхтекучести. "Стандартные" кривые остывания и кривые для сверхтекучих НЗ с учетом излучения нейтрино при куперовском спаривании нейтронов занимают промежуточное положение. ("Стандартное" остывание без прямых урка-процессов и без сверхтекучести, по-видимому, никогда не реализуется в природе и представляет теперь только исторический интерес, см., например, Каминкер и др. 2002).

Более подробное описание остывания НЗ можно найти в (Яковлев и др. 1999, Шааб и др. 1999, Паж и др. 2000, Яковлев и др. 2001).

Кроме перечисленных выше процессов на ход остывания НЗ существенное влияние могут оказать свойства внешних слоев НЗ (их атмосфер, см. раздел 2.2.2) и наличие дополнительных источников энергии. Сейчас известны три ситуации, когда поверхность или внешние слои НЗ дополнительно нагреваются. Это происходит в полярных шапках радиопульсаров, при аккреции и из-за распада магнитного поля в магнитарах. Рассмотрим их последовательно.

Стандартные модели пульсаров (Голдрайх, Джулиан 1969; Бескин и др. 1993) предсказывают, что в небольших областях вокруг магнитных полюсов электрическое поле, возникающее из-за вращения наклонного ротатора, будет вырывать из поверхности, а затем эффективно ускорять электроны. Через небольшое время ускоренные электроны начинают рождать $e^{\pm}$–пары. Вторичные электроны также начинают удаляться от НЗ, а позитроны движутся в обратном направлении и, так как их движение происходит практически вдоль силовых линий магнитного поля, то они попадают в области вблизи магнитных полюсов ("полярные магнитные шапки") и прогревают их до температуры $T \sim 10^6 - 10^7$ К. Именно они обуславливают пульсирующую часть теплового излучения, наблюдаемого у некоторых молодых и близких радиопульсаров (например у пульсара в Крабе). Этот механизм действует пока НЗ

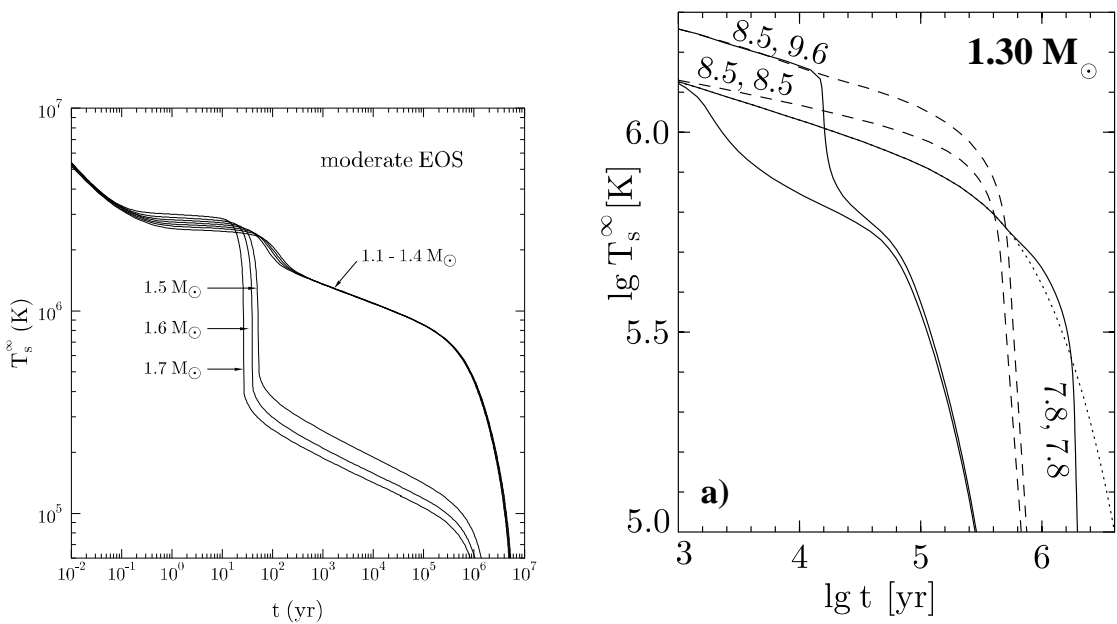

Рис. 1: Рисунки из обзоров Яковлева и др. (2001, 1999), иллюстрирующие различные типы остывания НЗ. На левом рисунке показаны стандартное (верхняя группа кривых для малых масс) и ускоренное (нижняя группа) остывание без сверхтекучести. На правом — стандартное (пунктир) и ускоренное (за счет куперовского спаривания нейтроны — сплошная линия) остывание сверхтекучих нейтронных звезд.

проявляет себя как радиопульсар, т.е. пока

$$\frac{B_{12}}{p^2} > 0.2,$$

где $B_{12} \equiv B/10^{12}$ Гс — магнитное поле на поверхности НЗ, а $p$ — период пульсара в секундах (Рудерман, Сазерленд 1975).

**Аккреция.** Условия наступления этой стадии описаны в параграфе 2.2.6 Для "стандартной" НЗ (радиопульсара) с $B \simeq 10^{12}$ Гс, $p_0 \simeq 0.01$–$0.1$ с, с медленным ($\tau > 10^8$ лет) затуханием магнитного поля аккреция из межзвездной среды (МЗС) начнется спустя не менее $10^9$ лет после ее рождения. В первую очередь это время зависит от пространственной скорости НЗ (см. параграф 2.2.6). Аккреция на молодые ($\sim 10^6$ лет) НЗ возможна при больших $p_0$ и/или меньших $B$. Однако, даже аккреция на старую НЗ ($t > 10^7$ лет) быстро прогревает ее кору и может сделать звезду видимой. При аккреции на НЗ со слабым полем вещество падает на всю поверхность звезды. Типичная температура одиночной аккрецирующей НЗ лежит в интервале 0.03-0.3 кэВ.

Для НЗ с более сильными полями аккрецирующее вещество будет двигаться вдоль магнитных силовых линий и выпадать на магнитные полюса НЗ. В этом случае температура излучения будет более высокой из-за меньшей площади излучающей области.

Аккреция на одиночные НЗ возможна не только из МЗС, но и из околозвездного диска, образующегося из остатков оболочки сверхновой (см. ниже). В этом случае темп аккреции может быть достаточно велик.

Следует также отметить, что наличие аккрецированной оболочки весьма сильно влияет на кривые остывания. Через оболочку из легких элементов тепло переносится легче, чем через железную, такая оболочка увеличивает тепловую светимость НЗ на нейтринной стадии остывания и ускоряет высвечивание тепловой энергии на фотонной стадии (см., Шабрие и др. 1997, Потехин и др. 1997).

**Сильные магнитные поля.** Магнитное поле, превосходящее $\sim 10^{11}$ Гс, может существенно повлиять на темп охлаждения НЗ. Это связано в первую очередь с изменением теплопроводности коры НЗ, теплоперенос в которой определяется электронами. Теплопроводность в коре НЗ при произвольном магнитном поле была изучена в работе Потехина (1999), а влияние магнитного поля на тепловую структуру оболочек и кривые остывания НЗ — в работе (Потехин, Яковлев 2001). Магнитное поле облегчает теплоперенос вблизи магнитных полюсов и затрудняет его вблизи экватора. Эти два процесса конкурируют, из-за чего при умеренно сильных полях (как у радиопульсаров) общая тепловая прозрачность оболочки НЗ уменьшается, а при сверхсильных полях (у магнитаров) она увеличивается.

Теория магнитаров и их сверхсильных полей находится на начальном этапе развития, поэтому здесь будет достаточно простых оценок (см. также Хейл и Хернквист 1997а,б). Магнитар со средним полем $\langle B \rangle$ внутри НЗ может обеспечить за счет его распада светимость $L$ в течение интервала времени

$$\tau_{\text{магн.}} = \frac{\langle B^2 \rangle}{8\pi} \cdot \frac{4\pi}{3} R_{\text{NS}}^3 \cdot \frac{1}{L} = \frac{\langle B^2 \rangle R_{\text{NS}}^3}{6L}.$$

Откуда видно, что при начальном поле $\langle B \rangle = 10^{15}$ Гс распад поля обеспечивает светимость $L = 10^{31}$–$10^{32}$ эрг/с на протяжении $\sim 10^6$ лет и может сказаться на поздних этапах остывания НЗ. Такую же светимость может обеспечить магнитар с $\langle B \rangle = 10^{14}$ Гс, но уже только в течение $\sim 10^4$ лет, что во-первых снизит количество наблюдаемых объектов (т.к. нам требуется попасть в узкий, $\sim 1\%$, интервал времени жизни объекта), а во-вторых потребует объяснения для позднего начала подобного затухания.

Однако, для многих молодых НЗ должно выполняться условие $B < 3 \cdot 10^{12}$ Гс и распад поля (по наблюдениям радиопульсаров) несущественен на ранних стадиях эволюции многих НЗ. Поэтому представляется важным рассмотреть охлаждение НЗ без усложняющего влияния магнитного поля, а также эффектов нагрева из-за внутреннего трения при замедлении вращения НЗ.

При расчетах остывания НЗ необходимо учитывать эффекты общей теории относительности, которые хотя и не меняют картину остывания качественно, существенно сказываются на численных результатах. Детальное обсуждение эффектов ОТО при остывании НЗ можно найти, например, в работе Паж и др. (2000).

### 2.2.2 Расчеты спектров НЗ с учетом атмосфер различного состава

Геометрически тонкий (несколько миллиметров, самое большее несколько сантиметров) наружный слой вещества коры НЗ существенным образом влияет на спектр испускаемого звездой излучения. Интересно отметить, что в то время как вопросы переноса излучения, структуры атмосфер и формирования спектров у обычных звезд изучаются многие десятилетия (Чандрасекар 1953; Соболев 1985) и в послевоенное время составляли основное содержание астрофизики, изучение атмосфер НЗ началось совсем недавно. Первая работа была выполнена Романи в 1987. Затем разными авторами были выполнены несколько циклов работ по моделированию атмосфер НЗ (Шибанов и др. 1992, Завлин и др. 1996, Раджагопал, Романи 1996 и другие). Обзор этих работ см. в (Вентура, Потехин 2001), а результаты недавних расчетов в (Хо, Лай 2001).

Моделирование атмосфер НЗ имеет свои особенности, как упрощающие, так и усложняющие расчеты, по сравнению с обычными звездами. К первым относятся крайне малая геометрическая толща атмосфер НЗ, позволяющая всегда пользоваться только плоско-параллельной геометрией, и предположение о гидростатическом равновесии. С высокой точностью выполняются также ионизационное и локальное термодинамическое равновесия. Ко второй группе факторов относятся худшая изученность рентгеновских спектров атомов по сравнению с оптическими, неопределенность химического состава атмосфер и наличие у НЗ сильного магнитного поля (от $10^8$ Гс до, возможно, $10^{15}$ Гс).

Рассмотрим два последние фактора. В работах разных авторов делались различные различных предположения о химическом составе атмосфер НЗ: чисто водородная, гелиевая и железная атмосферы, смесь с солнечным химсоставом и так называемый "кремниевый пепел" ("Si-ashes", вещество такого состава может выпадать на поверхность НЗ при аккреции самых внутренних слоев оболочки сверхновой). Конечно, всегда проводилось сравнение

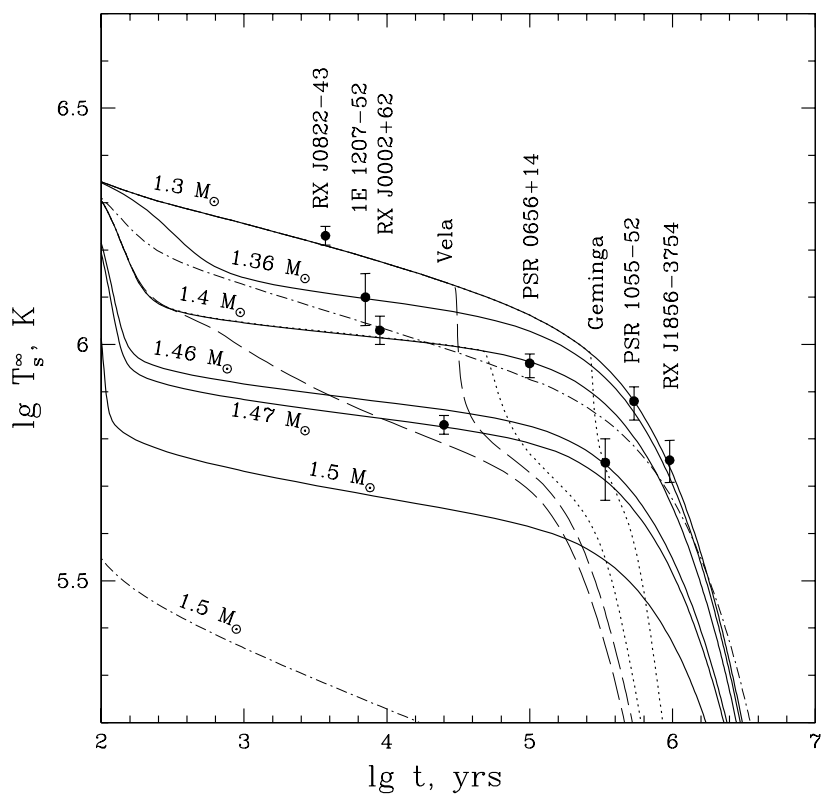

Рис. 2: Рисунок из работы Каминкер и др. (2001). Наблюдательные измерения поверхностной температуры восьми НЗ (показаны с ошибками измерений) из табл. 3 в сравнении с теоретическими кривыми остывания НЗ с протонной и нейтронной сверхтекучестью. Все кривые (за исключением штрих-пунктирных) построены для одной и той же протонной сверхтекучести. Сплошные линии показывают остывание нейтронных звезд с различной массой без нейтронной сверхтекучести. Штриховые и пунктирные линии соответствуют нейтронным звездам с массами $M = 1.3$ и $1.4\,M_\odot$, обладающими нейтронной сверхтекучестью. Штрих-пунктирными линиями показаны несверхтекучие НЗ с массами $1.3$ и $1.5\,M_\odot$. Сравнение теоретических кривых с наблюдениями позволяет жестко ограничить некоторые параметры НЗ.

с чернотельным планковским спектром. Наличие любой рассеивающей атмосферы делает спектр испускаемого излучения более жестким, что приводит к снижению определяемой по форме спектра температуры. Однако, результаты получаемые для тяжелых элементов и солнечного химсостава отличались от чернотельного спектра не более, чем на 20–30%. Сильные отличия (в 10 раз по потоку и в 2–3 раза по температуре) показывают только водородная и гелиевая атмосферы. При достаточно высоких температурах ($T > 10^7$ K) и/или плотностях ($\rho > 10^8$ г/см$^3$) водород на поверхности НЗ достаточно быстро выгорает в термоядерных или пикноядерных реакциях. В этом случае для поддержания атмосферы необходима хотя бы небольшая аккреция. Гелий на поверхности НЗ выгорает существенно медленнее, с другой стороны, присутствие в атмосфере даже незначительно количества CNO-элементов заметно ускоряет выгорание. Таким образом для моделирования большинства случаев обычно достаточно двух моделей: водородной атмосферы и черного тела.

Сильное магнитное поле ($B > 10^{11}$ Гс) оказывает на атмосферы НЗ очень существенное и сложное по своему характеру воздействие — перенос излучения становится анизотропным. Особенно сильным оказывается влияние магнитных полей на спектр выходящего излучения. В настоящее время в серии работ группы ФТИ им.А.Ф.Иоффе (Шибанов и др. 1992, 1995а,б, Павлов и др. 1995, Завлин и др. 1996, Потехин, Павлов 1997, Потехин и др. 1998) были построены спектры атмосфер, состоящих из элементов тяжелее гелия для полей до

Таблица 3: Экспериментально определенные поверхностные температуры восьми изолированных нейтронных звезд умеренного возраста. Таблица взята из работы Каминкер и др. (2002).

| Объект | lg $t$ [лет] | lg $T_s^\infty$ [K] | Модель[a] | Ссылки |
|---|---|---|---|---|
| RX J0822–43 | 3.57 | $6.23^{+0.02}_{-0.02}$ | H | Завлин и др. (1999) |
| 1E 1207–52 | 3.85 | $6.10^{+0.05}_{-0.06}$ | H | Завлин и др. (1998) |
| RX J0002+62 | 3.95[b] | $6.03^{+0.03}_{-0.03}$ | H | Завлин, Павлов (1999) |
| PSR 0833–45 (Vela) | 4.4[c] | $5.83^{+0.02}_{-0.02}$ | H | Павлов и др. (2001б) |
| PSR 0656+14 | 5.00 | $5.96^{+0.02}_{-0.03}$ | bb | Поссенти и др. (1996) |
| PSR 0633+1748 (Geminga) | 5.53 | $5.75^{+0.05}_{-0.08}$ | bb | Гальперн, Ванг (1997) |
| PSR 1055–52 | 5.73 | $5.88^{+0.03}_{-0.04}$ | bb | Огельман (1995) |
| RX J1856–3754 | 5.95 | $5.72^{+0.05}_{-0.06}$ | [d] | Понс и др. (2001) |
|  |  | $5.851 \pm 0.002$ | bb | Дрейк и др. (2002) |
| PSR J0205+6449 (3C58) | 2.91[e] | $< 6.05$ | все | Слэйн и др. (2002) |

[a] Температура определялась либо для модели водородной атмосферы (H), либо в предположение чернотельного спектра излучения (bb)

[b] Средний возраст взят согласно Крейгу и др. (1997).

[c] Согласно Лайну и др. (1996).

[d] Оценка температуры получена для аналитической модели Si-атмосферы (Понс и др. 2001).

[e] Пульсар отождествляется с исторической сверхновой 1181 года.

$10^{14}$ Гс, а для водородных атмосфер — до $10^{11}$ Гс (см. также Раджагопал и др. 1997). В работе (Потехин, Яковлев 2001) была сделана попытка рассмотреть остывание и атмосферы НЗ при полях до $10^{16}$ Гс. Эти модели уже используются при интерпретации спектров восьми НЗ, от которых зарегистрировано тепловое излучение (Завлин и др. 1996, Завлин, Павлов 1998, Гансик и др. 2002).

Спектры аккрецирующих одиночных НЗ изучались начиная с конца 60-х годов (Зельдович, Шакура 1969, Шварцман 1970в). В 90-е гг. много результатов было получено итальянской группой (Туролла и др. 1994, Зампьери и др. 1995).

В последние годы благодаря наличию на борту спутников XMM и Чандра спектрографов высокого (для рентгеновского диапазона) разрешения удается получать богатый наблюдательный материал для непосредственной проверки расчетов (см. например недавнюю работу Маршалл, Шульц 2002).

### 2.2.3 Процессы в магнитосферах

Теории магнитосфер для различных стадий эволюции НЗ разработаны на сегодняшний день в различной мере. Для стадий эжекции (E) и геротатора (G) есть достаточно детальные и более-менее самосогласованные модели магнитосфер. В первом случае их разработка была направлена на объяснение свойств радиопульсаров (Голдрайх, Джулиан 1969; Майкель 1991; Бескин и др. 1993; Муслимов, Цыган, 1990, 1992, Бескин 1990, Цыган 1993, Боговалов 1999, 2001, Любарский 1995), во втором — на описание взаимодействия магнитосферы Земли с Солнечным ветром (например, Жигулев, Ромашевский 1959). Для стадий аккреции и пропеллера получены только существенно более простые результаты при дополнительных модельных предположениях (см., например, Липунов 1987). Следует заметить, что в то время как в магнитосферах эжекторов и георотаторов присутствуют как замкнутые, так и уходящие на бесконечность силовые линии, магнитосферы пропеллеров и аккреторов могут быть полностью замкнуты (при сферической аккреции).

Как неоднократно отмечалось в обзорах и докладах В.С Бескина, в последние годы снизилась активность авторов, исследующих магнитосферы радиопульсаров. Однако, наметилась новая область исследований — магнитосферы *магнитаров* — особенно сильно замагниченных НЗ (Д.Г. Яковлевым было замечено, что уместнее именно такое написание данного термина, несмотря на устоявшееся воспроизведение англоязычного варианта). Исследования магнитаров особенно актуальны в приложении к МПГ (см., например, Томпсон, Дункан 1996).

Сильные магнитные поля, которыми обладают обычные НЗ, и, тем более, магнитары, существенным образом влияют на элементарные физические процессы вблизи поверхностей этих объектов. Отметим ряд критических значений напряженности магнитного поля:

- НЗ с полем порядка $10^8$ Гс и менее можно рассматривать как незамагниченные, поскольку характерное время замедления вращения такого объекта превышает возраст Вселенной, а при аккреции магнитное поле прижимается к поверхности НЗ потоком падающего вещества ($R_A < R_{NS}$) и перестает оказывать какое-либо влияние (здесь ($R_A-$ альвеновский радиус).

- При $B > \alpha^2 B_{\rm Sh} = 2.4 \cdot 10^9$ Гс (здесь $\alpha \equiv e^2/\hbar c \simeq 1/137$ — постоянная тонкой структуры, а $B_{\rm Sh} \equiv m_e^2 c^3/(\hbar e) = 4.4 \cdot 10^{13}$ Гс — Швингеровское поле) гирорадиус электрона становится меньше радиуса Боровской орбиты в атоме водорода. Более сильные поля оказывают сильное влияние на структуру атомов, которые не ионизуются полностью даже на поверхности магнитаров (Томпсон, Дункан 1995; Хейл, Хернквист 1997a). В первую очередь эти изменения сказываются на свойствах атмосфер НЗ (см. параграф 2.2.2).

  Ближе к верхней границе интервала важными становятся анизотропия теплопроводности в коре НЗ (см. 2.2.1) и рассеяния излучения в атмосфере (см. 2.2.2).

- В интервале $B_{\rm Sh} = 4.4 \cdot 10^{13} < B \lesssim 10^{18}$ Гс начинает происходить еще ряд интересных процессов. Энергия первого уровня Ландау начинает превышать энергию покоя электрона. Становится существенными реакции расщепления фотона ($\gamma \to 2\gamma$), однофотонного рождения электрон-позитронной пары или ее аннигиляции ($\gamma \longleftrightarrow e^+ + e^-$). Причем сечение последнего процесса для фотонов с различной поляризацией в сильных полях ($\gg B_{\rm Sh}$) существенно различается (Томпсон, Дункан 1995). Без магнитного поля эти процессы невозможны в принципе, а в слабых магнитных полях ($B \ll B_{\rm Sh}$) они происходят только для очень энергичных фотонов ($h\nu \gg m_e c^2$). Анизотропия атмосферного рассеяния и теплопроводности в коре НЗ еще более усиливается.

  Верхняя граница интервала ($\sim 10^{18}$ Гс) определяется равенством энергии магнитного поля и гравитационной энергии связи НЗ. Более сильные стационарные магнитные поля по-видимому не могут существовать на НЗ.

Подробнее о новых результатах физики в сверхсильных магнитных полях см., например, Томпсон (2000), Михеев (2000) или Дункан (2000) (а также материалы конференции "Сильные магнитные поля в нейтринной астрофизике" (2000) и старый, но очень хороший обзор Павлова и Гнедина (1983), посвященный поляризации вакуума и элементарным процессам в сильных магнитных полях).

Отметим дискуссию по поводу отсутствия пульсарного радиоизлучения магнитаров (см. работы Усова и Мелроуза 1996 и Баринг и Хардинг 1995 и др.). Расщепление фотонов в поле магнитара конкурирует с рождением пар и приводит к быстрому снижению энергии жестких фотонов. Однако не ясно достаточно ли эффективен этот процесс, чтобы объяснить полное отсутствие пульсирующего радиоизлучения у магнитаров. Отметим также работы (Томпсон и др. 2001, Лютиков и др. 2001), посвященные генерации жесткого излучения — мягких гамма-всплесков — в магнитосферах сильнозамагниченных НЗ, и работу Бастуркова с соавторами (2002), в которой периодов магнитаров объясняются вращательными (нерадиальными) колебаниями НЗ.

Современные спутники (Чандра, Ньютон) позволяют надеяться на открытие деталей в спектрах АРП и МПГ. Поэтому детальные расчеты рентгеновских спектров становятся очень актуальной задачей. Наиболее интересные результаты получены Дзане и др. (2001) — ими предсказана возможность обнаружения с помощью современных детекторов протонной циклотронной линии, которая у магнитаров с полями $10^{14}$–$10^{15}$ Гс попадает в "стандартный" рентгеновский диапазон 2–10 кэВ: $\sim 0.63\, B/10^{14}$ Гс кэВ (см. также недавние расчеты Озела 2002).

### 2.2.4 Расчеты затухания магнитного поля НЗ

Молодые НЗ (радиопульсары, НЗ в массивных двойных системах и т.д.) имеют сильные магнитные поля. Старые НЗ (миллисекундные пульсары) имеют слабые поля. Таким образом логично предположить наличие механизма затухания поля. Затухание магнитного поля становится все более стандартным предположением при рассмотрении эволюции НЗ и описании их свойств. Однако, вопрос о механизме диссипации поля остается открытым. С этим связано появление в последние годы большого числа работ по этой теме (см. Митра и др. 2000, Таурис и Конар 2001 и др.).

Прежде всего необходимо понять, сконцентрировано ли магнитное поле в коре НЗ, или же пронизывает и ее ядро, а также какую роль играет аккреция. Механизмы, ответственные за эволюция поля в коре и ядре НЗ, различны. Наиболее полное исследование распада поля в коре (без учета эффектов ОТО) было проведено в работе (Урпин, Коненков 1997) (см. также Урпин и Муслимов 1992). Расчеты затухания поля, сосредоточенного в коре, с учетом эффектов ОТО приведены в (Гепперт и др. 2000; Паж и др. 2000).

Основные результаты расчета распада дипольного магнитного поля в коре одиночной НЗ таковы. Диссипация магнитного поля оказывается тесным образом связанной с тепловой эволюцией НЗ. Для *стандартного* остывания, при котором нейтринная светимость НЗ определяется в основном модифицированными урка-процессами (Петик, 1992), за первый миллион лет поле распадается в 2–1000 раз в зависимости от начальной глубины залегания и уравнения состояния в ядре звезды (Урпин, Коненков 1997, см. рис. 4). По мере остывания НЗ проводимость увеличивается, и распад поля замедляется. Скорость распада на поздней стадии зависит от примесной проводимости $\sigma_{imp}$ (см. рис. 4). Например, при типичном значении $\sigma_{imp}$, принятом в (Урпин, Коненков 1997), поле практически не уменьшается за последующие $10^8$ лет. Однако как только магнитное поле продиффундирует через всю кору и достигнет сверхпроводящего ядра (за $2 \cdot 10^9$ лет при той же $\sigma_{imp}$), распад становится экспоненциальным.

Аккреция оказывает влияние на эволюцию поля. Во-первых, она нагревает кору нейтронной звезды (Ждуник и др., 1992), уменьшая тем самым проводимость. Во-вторых, возникает поток вещества, направленный к центру звезды, который стремится перенести поле в более глубокие слои. Как показывают расчеты (Гепперт и др. 1996), аккреция с темпом $\dot M < 10^{-14} M_\odot$/год незначительно ускоряет распад поля. Таким образом для одиночных НЗ этим эффектом можно пренебречь.

Возможен механизм, в котором поле из ядра НЗ "выталкивается" в кору за счет вращения или архимедовой силы (см. Муслимов и Цыган 1985), и там уже затухает за счет омических потерь. Недавно вычисления для этого механизма были проведены Коненковым и Геппертом (2000, 2001а,б).

В работе (Колпи и др. 2000) рассмотрены три механизма распада, применимые в случае сверхсильных полей (магнитары). Авторы делают вывод, что с точки зрения наблюдений наиболее привлекательным является механизм Холловского каскада (Hall cascade).

С точки зрения эволюции одиночных НЗ распад поля может давать разные эффекты. Для некоторого набора параметров распад может уменьшить число НЗ на стадии аккреции (Колпи и др. 1998, Ливио и др. 1998), для другого набора — может увеличить это число (см. Попов, Прохоров 2000 и рис. 5). В первом случае из-за распада НЗ "застынет" на стадии пропеллера. Во втором (быстрый распад до малых полей) — исчезновение поля приведет к исчезновению барьера, и вещество сможет беспрепятственно выпадать на поверхность НЗ,

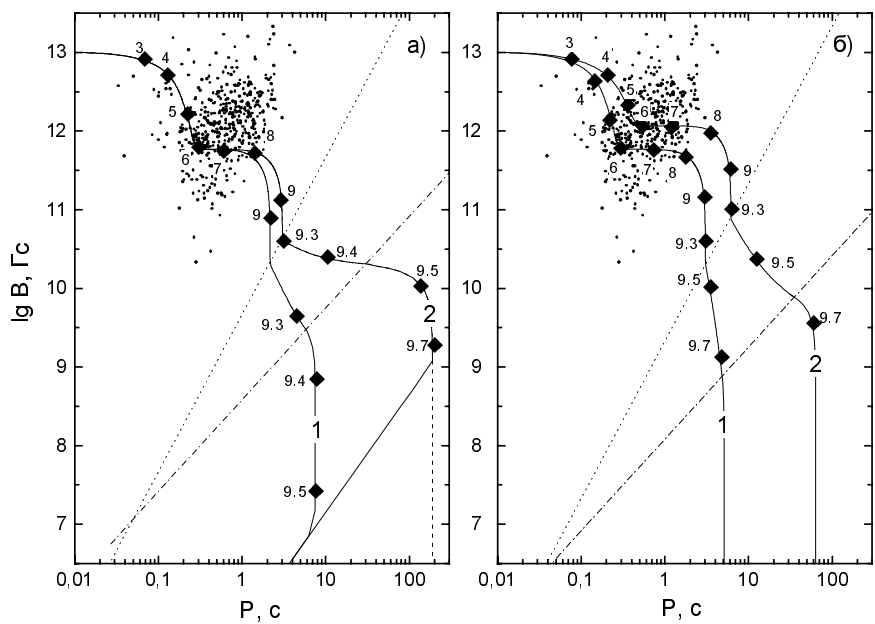

Рис. 3: Эволюционные треки НЗ для темпа аккреции $\dot M = 10^{-15} M_\odot$год$^{-1}$ (a) и $\dot M = 10^{-16} M_\odot$ год$^{-1}$ (b) (Коненков, Попов 1997). Целью вышеуказанной работы было воспроизведение наблюдаемых параметров источника RX J0720.4-3125 в модели аккрецирующей НЗ с затухшим полем. Пунктирные линии соответствуют $p = P_E$; штрих-пунктирные — $p = P_A$. Числа возле отметок на треках отмечают логарифм возраста НЗ в годах. Точками показаны радиопульсары (Тейлор и др. 1993).

начнется аккреция.

Затухание магнитного поля активно используется в популяционном синтезе радиопульсаров и НЗ других типов (см. ниже).

### 2.2.5 Аккреция на одиночные НЗ из межзвездной среды и из околозвездных остаточных дисков

Классическая теория аккреции уходит корнями еще в 30-40-е гг. (Бонди и Хойл 1944, Хойл и Литтлтон 1939, см. также ссылки на ранние работы в книге Горбацкого 1977). Аккрецию на одиночные НЗ начали рассматривать еще в самом начале 70-х гг. (см. Шварцман 1970в, Острайкер и др. 1970), когда стало очевидным, что аккрецирующие НЗ являются источниками рентгеновского излучения.

Физика аккреции на одиночные объекты несколько отличается от аккреции в двойных системах. Во-первых, отсутствует орбитальный момент. Во-вторых, чаще всего (при аккреции из МЗС) темп аккреции невелик (см. Тревес и др. 1993).

Для астрофизических приложений ключевым вопросом является определение темпа аккреции, $\dot M$. Все рассмотрение проводится для столкновительной среды, что всегда выполняется в случае аккреции на НЗ из МЗС.

Современные работы направлены на учет различных эффектов, которые могут изменять (обычно уменьшать) темп аккреции по-сравнению с классическими результатами Бонди и др.

Очевидно, что темп аккреции можно записать в виде: $\dot M = \sigma \rho_\infty v_\infty$. Определим сечение $\sigma$. В случае сферической аккреции радиус гравитационного захвата будет равным:

$$R_G = 2GM/c_s^2, \qquad (1)$$

где $c_s$ — скорость звука в в МЗС вдали от НЗ. Таким образом $\dot M = \pi R_G^2 \rho_\infty c_s \propto c_s^{-3}$. Отметим сильную зависимость темпа аккреции от температуры $\dot M \propto T_\infty^{-3/2}$. Поэтому вопрос о прогреве МЗС излучением НЗ является исключительно важным. Остановить аккрецию прогрев

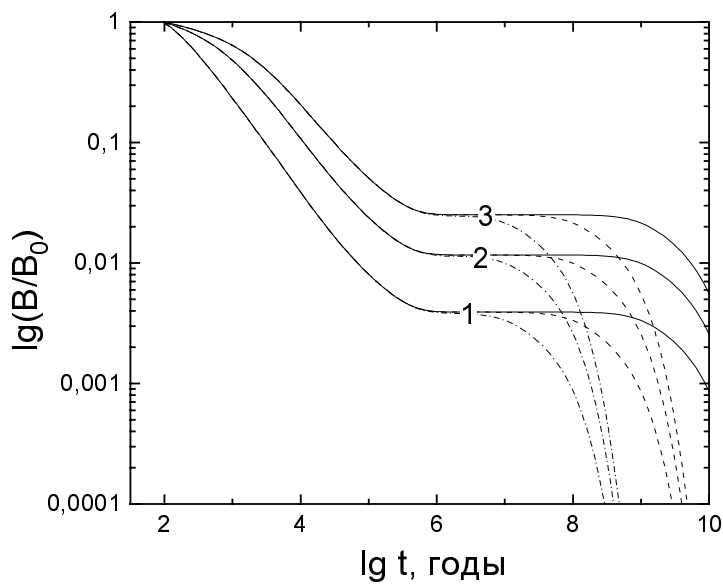

Рис. 4: Затухание магнитного поля НЗ (Коненков, Попов 1997). Кривые 1, 2, 3 соответствуют различным начальным глубинам залегания $10^{11}$, $10^{12}$ и $10^{13}$ г/см$^3$, соответственно. Сплошные кривые соответствуют Q = 0.001, штриховые — Q = 0.01, штрих-пунктирные — Q = 0.1. Q — параметр, характеризующий концентрацию и заряды примесей в коре НЗ.

не может (Бисноватый-Коган, Блинников 1980), однако может заметно уменьшить ее темп. Влияние прогрева будет рассмотрено ниже.

В случае цилиндрической аккреции ($v_\infty > c_s$) изменяется формула для радиуса гравитационного захвата:

$$R_G = 2GM/(c_s^2 + v_\infty^2). \qquad (2)$$

И таким образом для темпа аккреции имеем:

$$\dot{M} = k\pi \frac{(2GM)^2}{(c_s^2 + v_\infty^2)^{3/2}} \rho_\infty. \qquad (3)$$

Коэффициент пропорциональности, $k$, зависит от скорости НЗ. Грубо можно положить его равным единице. Точные аналитические решения в данном случае отсутствуют. Важным также оказывается учет влияния магнитосферы НЗ (Торопина и др. 2001).

Обратное влияние излучения аккрецирующего объекта на аккреционный поток не раз рассматривалось разными авторами, начиная с Шварцмана (1970а). В случае одиночной НЗ светимость далека от эддингтоновской, $L < 10^{-4} L_E$. Однако, как показано в работе Блаез и др. (1995), прогрев может уменьшить темп аккреции еще в несколько раз, что может многократно уменьшить число наблюдаемых одиночных НЗ.

За счет жесткого излучения движущейся НЗ вокруг нее возникает кометообразная туманность размером порядка $10^{17}$ см для самых низкоскоростных НЗ при плотности порядка $10^{-24}$ г см$^{-3}$ (Блаез и др. 1995). При движении в среде с плотностью порядка $10^{-24}$ г см$^{-3}$ уменьшение темпа аккреции составляет от 30 раз при скорости порядка 20 км с$^{-1}$ до 3 раз при скорости 40 км с$^{-1}$, при скоростях > 60 км с$^{-1}$ отличия становятся несущественными. Также различие становится менее существенным при больших плотностях МЗС (о характеристиках МЗС см. книгу Бочкарева 1992).

Численное моделирование сферической и цилиндрической аккреции на НЗ проводилось неоднократно. Остановимся на недавних расчетах аккреции на НЗ с учетом магнитного поля (Торопин и др. 1999, Торопина и др. 2001) (результаты этой группы также доступны в Интернете по адресу **http://www.astro.cornell.edu/us-russia/**).

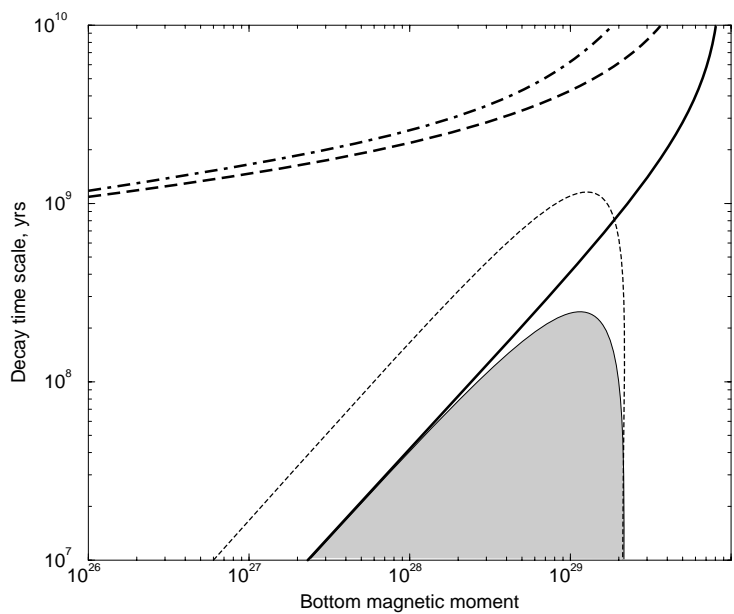

Рис. 5: Рисунок иллюстрирует области параметров, в зависимости от которых одиночная НЗ успевает выйти на стадию аккреции, или же остается на стадии Эжектора (или Пропеллера). По осям отложены характерный масштаб экспоненциального затухания магнитного поля, $t_d$, и минимальный магнитный момент, $\mu_b$ (ниже этой величины поле не распадается). В закрашенной области для НЗ с начальным магнитным моментом $\mu_0 = 10^{30}$ Гс см$^3$ время жизни на стадии эжекции, $t_E$, превосходит время жизни Галактики, $10^{10}$ лет. Таким образом из этой области параметров НЗ не может стать Аккретором. Штриховая линия соответствует условию $t_H = t_d \cdot \ln(\mu_0/\mu_b)$, где $t_H = 10^{10}$ лет, т.е. НЗ достигает минимального значения поля за Хаббловское время. Сплошная линия соответствует $P_E(\mu_b) = p(t = t_{cr})$, где $t_{cr} = t_d \cdot \ln(\mu_0/\mu_b)$, она определяет левую границу "запрещенной" области. Правая граница определяется минимальным начальным полем, при котором без распада НЗ успевает выйти на стадию аккреции. Эти линии и заштрихованная область нарисованы для $\mu_0 = 10^{30}$ Гс см$^3$. Штрих-пунктирная линия аналогична штриховой, но нарисована для $\mu_0 = 5 \cdot 10^{29}$ Гс см$^3$. Пунктир очерчивает область, аналогичную заштрихованной, для $\mu_0 = 5 \cdot 10^{29}$ Гс см$^3$ (Попов, Прохоров 2000).

В случае сферически-симметричной аккреции в расчетах было получено уменьшение темпа аккреции примерно в 2 раза по сравнению с формулой Бонди. Темп аккреции на магнитный диполь зависит от магнитного поля НЗ и от магнитной проницаемости среды: $\dot M_{dip} \propto \eta_m^{0.38}$, $\dot M_{dip} \propto (\dot M_B/\mu^2)$, где $\eta_m$ — магнитная проницаемость, а $\dot M_B$ — темп аккреции Бонди.

В случае цилиндрической аккреции показано, что при наличии магнитного поля темп аккреции уменьшается в несколько раз по сравнению с незамагниченной НЗ. Чем больше магнитное поле, тем меньше темп аккреции: $\dot M \propto B^{-1.3}$.

Таким образом современные исследования показывают, что темп аккреции Бонди является верхним пределом, редко реализующимся в природе.

В последние несколько лет появилась серия работ, посвященных остаточным (remnant) аккреционным дискам вокруг молодых НЗ (см. Ротшильд и др. 2001, Мену и др. 2001а, Чаттерье и др. 2000, Марсден и др. 2000 и ссылки там). Эти исследования связаны с гипотезой о том, что активность многих молодых радиотихих НЗ связана не с энергией магнитного поля и не с остаточным теплом, а с аккрецией. Темп аккреции является функцией времени, т.к. новое вещество в диск не поступает. Аккреционный диск образуется из вещества остатка взрыва, захваченного гравитационным полем образовавшегося компактного объекта (fall-back). В настоящее время процесс обратного выпадения вещества считается стандартным в моделях взрыва сверхновой. Ранее эволюция таких дисков в основном рассматривалась в связи с образованием планет около радиопульсаров.

Получены оценки темпа аккреции, времени существования диска, изменения периода вращения НЗ, а также рассчитаны спектральные характеристики излучения. Исследованы диски разного химического состава, поскольку вещество внутренних частей остатка сверхновой должно быть существенно обогащено тяжелыми элементами. Рассчитана устойчивость дисков на разных стадиях их эволюции (Мену и др. 2001а).

В данных моделях (см. Ротшильд и др. 2001) показано, что для получения нужного эффекта достаточно диска с массой $\sim 10^{-5} M_\odot$, хотя его начальная масса может быть существенно большей (до $\sim 0.01 M_\odot$). Типичный возраст НЗ оказывается порядка тысяч и десятков тысяч лет, что находится в соответствии с независимыми оценками (в первую очередь по возрасту остатка сверхновой). Вспышки МПГ объясняются процессами в коре НЗ. Спектр диска существенно отличается от дисков Шакуры-Сюняева из-за наличия пыли.

Во многих моделях (см. Чаттерье и др. 2000, Альпар 2001) НЗ находятся не на стадии аккреции, а на стадии Пропеллера. Были предложены сценарии, в которых аккреционный диск вносит вклад в увеличение периода радиопульсаров (Мену и др. 2001б, Альпар и др. 2001). В этом случае индекс торможения оказывается меньшим 3. Например, для пульсара в Крабе требуется диск с втокома вещества $3 \cdot 10^{13} - 10^{17}$ г/с. Для пульсара Vela требуется диск очень малой массы с сильным наклоном диск к оси вращения.

Подбором параметров, авторам удается построить "единую модель" эволюции одиночных НЗ, описывающую все наблюдаемые типы объектов (Альпар 2001). Однако, на данный момент в таких работах есть много непроработанных деталей, что приводит к тому, что модели выглядят несколько искусственными.

### 2.2.6 Расчеты эволюции периодов вращения НЗ

Из всех параметров НЗ наиболее точно измеряются периоды их вращения. Кроме этого измерения периодов являются модельно независимыми (в отличии от температуры поверхности, массы одиночных НЗ и т.д.). Поэтому необходимо иметь хорошее описание эволюции этого параметра.

Распределение начальных периодов НЗ неизвестно (см. ван дер Сваллоу, Ву 2001, Регимбо, де Фрейтас Пачеко 2001, Шевалье и Эммеринг 1986). Наблюдения молодых НЗ с известными периодами может дать важные ограничения (см. Готтхелф и др. 1999).

На начальные периоды НЗ существенное влияние могут оказывать т.н. r-моды, связанные с излучением гравитационных волн (Андерсон и др. 2000, Оуэн и др. 1998, о гравитационных волнах от НЗ см. также Брагинский 2000, Грищук и др. 2000 и Джиазотто и др. 1997, о неустойчивостях и колебаниях НЗ — Баструков и др. 1999, Линдблом 2001).

Разумно выделить четыре основные стадии эволюции НЗ: Эжектор, Пропеллер, Аккретор и Георотатор (см. Липунов 1987, Липунов и др. 1996). На стадии Эжектора поток электромагнитных волн и релятивистских частиц от НЗ "выдувает" окружающее вещество за пределы всех характерных радиусов. Типичными представителями Эжекторов являются радиопульсары. Однако, стадия радиопульсара заканчивается раньше стадии Эжектора (Аронс 2000, Чен и Рудерман 1993). Продолжительность стадии Эжектора при постоянном поле составляет:

$$t_E \approx 10^9 \mu_{30}^{-1} n^{1/2} v_{10}$$

На стадии Пропеллера аккреция невозможна из-за наличия быстровращающейся магнитосферы. Георотатором мы называем стадию, на которой радиус магнитосферы настолько велик, что вещество не захватывается НЗ гравитационно.

Конкретное состояние НЗ определяется соотношениями между четырьмя характеристическими радиусами: $R_l = c/\omega$ – радиусом светового цилиндра, $R_{st}$ – радиусом остановки (например альвеновский радиус, $R_A$, это частный случай радиуса остановки, см. Липунов 1987), $R_G = (2GM)/v^2$ – радиусом гравитационного захвата и $R_{co} = (GM/\omega^2)^{1/3}$ – радиусом коротации. Здесь $M$ – масса НЗ, $c$ – скорость света, $\omega$ – частота вращения, $v^2 = v_\infty^2 + c_s^2$, $c_s$ – скорость звука, $v_\infty$ – скорость НЗ относительно МЗС.

Соотношение между радиусами определяет два критических периода: $P_E$ и $P_A$, разделяющих различные стадии эволюции НЗ. Эти периоды могут быть оценены по формулам

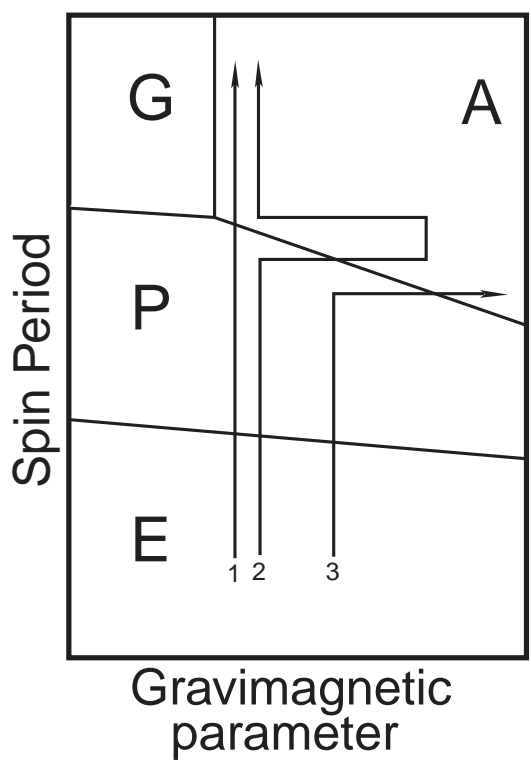

Рис. 6: $p-y$ диаграмма, позволяющая проиллюстрировать эволюцию одиночной НЗ (см. Липунов 1987, Липунов, Попов 1995). По горизонтальной оси отложен гравимагнитный параметр, $y = \dot M/\mu^2$. Схематично показаны три эволюционных трека. 1 — постоянные магнитное поле и внешние условия, 2 — пролет области плотного газа (например, молекулярного облака), 3 — трек с затуханием магнитного поля.

(Липунов 1987):

$$P_E = 2\pi \left(\frac{2\,k_t}{c^4}\right)^{1/4} \left(\frac{\mu^2}{v\dot M}\right)^{1/4}, \quad R_l < R_G, \tag{4}$$

$$P_A = 2^{5/14}\pi(GM)^{-5/7} \left(\frac{\mu^2}{\dot M}\right)^{3/7}, \quad R_A < R_G. \tag{5}$$

Здесь $\mu$ – магнитный дипольный момент, $\dot M \equiv \pi R_G^2 \rho v$ – темп аккреции, $\rho$ – плотность МЗС, $k_t$ – безразмерная константа порядка единицы.

Если $p < P_E$, то НЗ находится на стадии Эжектора; если $P_E < p < P_A$, мы имеем НЗ на стадии Пропеллера; наконец, если $p > P_A$ и $R_{st} < R_G$, то НЗ является Аккретором.

Когда $p > P_A$, но $R_{st} > R_G$ то аккреция невозможна, т.к. образуется геоподобная магнитосфера. Заметим, что замедление на стадии Георотатора подобно замедлению на стадии Пропеллера (см. ниже). Численно стадия Георотатора исследовалась в работах (Торопина и др. 2001, Романова и др. 2001). Также некоторые аспекты рассматривались в (Рутледж 2001).

На стадии Эжектора эволюция периода определяется потерями энергии вращения НЗ на излучение:

$$\dot p = \frac{8\pi^2 R^6}{3c^3 I} \cdot \frac{B^2(t)}{p}, \tag{6}$$

где $R$ – радиус НЗ, $I$ – момент инерции, $B = \mu/R^3$ – магнитное поле.

Существуют различные интерпретации этого замедления (см. Бескин и др. 1993). Однако, во всяком случае магнитодипольная формула хорошо описывает наблюдения (существу-ют, однако, работы, привлекающие дополнительное замедление, связанное с существованием остаточного аккреционного диска, см. пункт, посвященный аккреции).

На стадии Пропеллера НЗ замедляется из-за передачи углового момента окружающему веществу (Шварцман 1970в, Илларионов, Сюняев, 1975). Существует множество формул, описывающих замедление на стадии Пропеллера (см. Липунов 1987, Липунов, Попов 1995). Фактически все они сводятся к виду:

$$\frac{dI\omega}{dt} = -k_t \frac{\mu^2}{R_A^3}. \tag{7}$$

Множитель $k_t$ будет различным в разных моделях, в том числе он может зависеть от частоты вращения НЗ.

В работе (Липунов, Попов 1995) было сформулировано важное утверждение: для постоянного магнитного поля длительность стадии Эжектора при разумных параметрах всегда больше длительности стадии Пропеллера. В случае затухания поля это может быть не так. В недавних работах было показано, что распад магнитного поля может как увеличить количество НЗ на стадии Пропеллера (Колпи и др. 1998, Ливио и др. 1998), так и уменьшить его, если распад очень быстрый и идет до низких значений полей (Попов, Прохоров 2000).

На стадии аккретора на НЗ действуют два момента сил:

$$\frac{dI\omega}{dt} = K_{sd} + K_{turb}, \tag{8}$$

$$K_{sd} = -k_t \frac{\mu^2}{R_{co}^3}.$$

Здесь $K_{sd}$ – тормозящий момент сил, связанный с магнитным полем НЗ, а $K_{turb}$ – момент сил, возникающий из-за того, что МЗС может быть сильно турбулизована. $K_{turb}$ действует случайно, и может как ускорять, так и замедлять НЗ (см. Липунов, Попов 1995).

Отметим, что в приведенной выше формуле мы можем как переоценивать, так и недооценивать замедляющий момент, т.к. детали передачи момента внешней среде неясны. Если указанная выше формула применима, то аккреция должна быть существенно дозвуковой, а значит темп аккреции на НЗ будет ниже определяемого по формуле Бонди. При эффективной аккреции, соответствующей формуле Бонди, темп уноса углового момента должен быть меньше.

Изменение периода аккрецирующей НЗ связано со взаимодействием с турбулизованной МЗС. Это привносит свою специфику в задачу об эволюции периода. Если принять гипотезу об ускорении нейтронной звезды в турбулизованной МЗС (Липунов, Попов 1995), то возникает новый характерный период, $P_{eq}$. Он определяется условием квазиравновесия замедления (оно определяется как торможением за счет магнитного поля НЗ, так и аккрецируемым моментом) и ускорения (определяется аккрецируемым моментом).

В реальной ситуации (Прохоров и др. 2002) квазиравновесие не достигается (рис. 8). Для определения распределения НЗ по периодам надо решать дифференциальное уравнение, используя реалистичные распределения НЗ по скоростям и магнитным полям.

Расчеты для постоянного магнитного поля и максвелловского распределения НЗ по скоростям показали наличие широкого максимума в распределении вблизи периодов $10^6$–$10^7$ секунд (Прохоров и др. 2002). Распределение периодов для разных параметров показано на рис.9. Таким образом в такой модели следует ожидать отсутствия наблюдений периодов вращения аккрецирующих НЗ (типичное время наблюдения на рентгеновском спутнике порядка $10^4 - 10^5$ секунд).

Однако, в случае распада поля ситуация будет совсем иной (Ванг 1997, Коненков, Попов 1997). Период "застынет", запомнив значение, соответствующее начальному полю (см. рис. 3), и можно ожидать появления аккрецирующих НЗ с периодами порядка 10 секунд.

Качественная картина эволюции периода для случая постоянного поля показана на рис. 7.

Распределение периодов АРП и МПГ исследовалось в работе Псалтиса и Миллера (2002). В этой статье авторам не удалось получить серьезных ограничений на значение начального периода, т.к. показатель торможения для АРП и МПГ известен очень плохо. Единственным

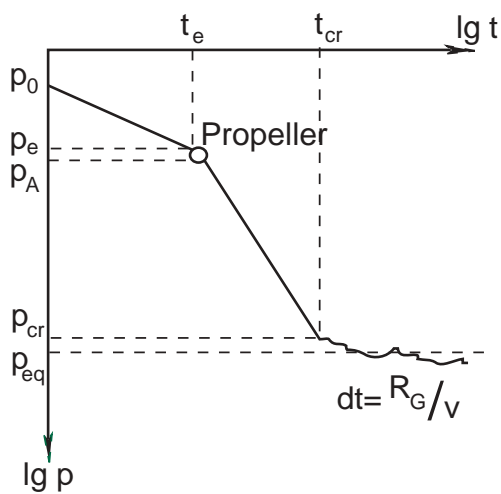

Рис. 7: Эволюция одиночной НЗ в турбулизованной МЗС. После замедления на стадии Эжектора, когда период достигает $p = P_E$, и короткой стадии Пропеллера, показанной на рисунке кружком, НЗ попадает на стадию Аккретора ($p > P_A$). В начале этой стадии торможение за счет магнитного поля оказывается более существенным, чем изменение момента за счет аккреции турбулизованной среды. Затем в момент $t = t_{cr}$ эти два эффекта сравниваются, и период начинает флуктуировать. На этой стадии типичным периодом является $P_{eq}$, однако отклонения могут быть очень значительными. Типичный временной масштаб флуктуаций $dt = R_G/v$ (подробнее см., Попов 2001).

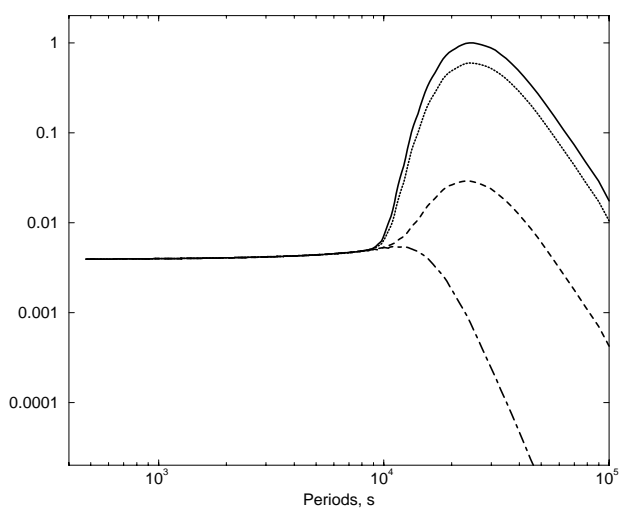

Рис. 8: Эволюция распределения периодов вращения одиночных аккрецирующих НЗ (Прохоров и др. 2002). Параметры задачи: $\mu = 10^{30}$ Гс·см$^3$, $n = 1$ см$^{-3}$, $v_{NS} = 10$ км/с. Кривые относительных плотностей распределения по периодам нарисованы для четырех моментов времени от $1.72 \cdot 10^9$ лет до $9.8 \cdot 10^9$ лет с момента рождения НЗ. Для выбранных параметров $t_A \simeq 1.7 \cdot 10^9$ лет. Видно, что требуется значительное время для достижения квазистационарного распределения (для некоторых наборов параметров время достижения такого состояния больше Хаббловского). НЗ пересекает горизонтальную часть распределения от $\simeq 10^2$ с до $10^4$ с примерно за $\sim 6 \cdot 10^7$ лет. Кривые нормированы на единицу в максимуме самой высокой кривой.

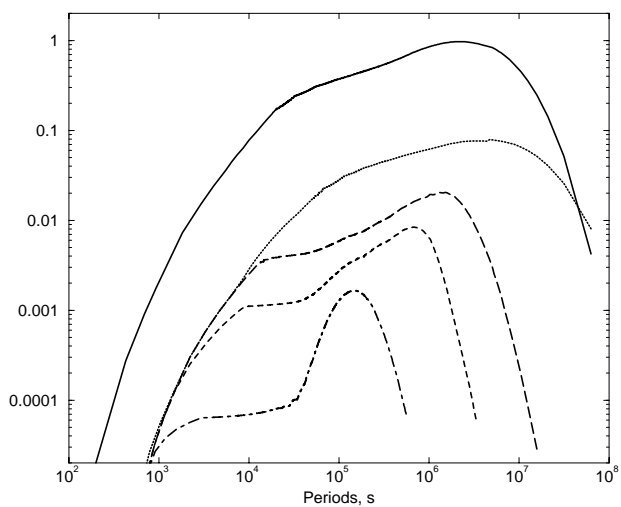

Рис. 9: Распределение периодов вращения одиночных аккрецирующих НЗ (Прохоров и др. 2002). В этих расчетах распределение НЗ по скоростям было максвелловским с дисперсией 140 км/с, магнитные поля были распределены в соответствии с данными для радиопульсаров. Верхняя кривая показывает относительное распределение НЗ по периодам для концентрации МЗС $n = 1$ см$^{-3}$. Вторая (пунктирная) — для $n = 0.1$ см$^{-3}$. Другие кривые рассчитаны для низкоскоростных НЗ ($v < 60$, 30, 15 км/с соответственно) при $n = 0.1$ см$^{-3}$. Кривые нормированы на единицу в максимуме самой верхней.

положительным (хотя и очевидным) результатом является необходимость "выключения" АРП и МПГ на периодах близких к максимальным из наблюдающихся (порядка 12 с).

В расчетах магнитовращательной эволюции НЗ существует еще много нерешенных вопросов. Российские научные группы активно участвуют в исследовании этих процессов и их вклад можно считать определяющим.

### 2.2.7 Роль межзвездного поглощения

Для многих исследований, рассматриваемых в данном обзоре, важным является учет межзвездного поглощения в рентгеновском диапазоне (см., например, Попов и др. 2000б). Межзвездное поглощение велико в мягком рентгеновском диапазоне ($\sim 0.1 - 1$ эВ) (см. Вилмс и др. 2000). Поэтому наблюдения рентгеновских источников на различных энергиях демонстрируют нам выборки объектов с разной селекцией по расстоянию. Из вышеперечисленных типов источников этот эффект наиболее существенен для слабых рентгеновских источников в диске Галактики. Объекты этого типа мы не можем наблюдать уже на расстояниях порядка 1 кпк.

Некоторые из рассматриваемых нами источников имеют максимумы излучения вблизи 0.1 кэВ. Спутники, с которых производят наблюдения этих объектов, работают в диапазонах энергий $\sim 0.1$–1 кэВ, некоторые выше, до $\sim 10$ кэВ. В этом интервале энергий поглощение рентгеновского излучения в межзвездной среде достаточно велико и обязательно должно учитываться.

В самом общем виде поглощение может быть описано формулой:

$$I = I_0 \exp\left(-\sigma_{\text{ISM}}(E, Z) \cdot N_{\text{H}}(\mathbf{n}, r)\right). \tag{9}$$

Здесь $I$ и $I_0$ — интенсивности излучения на детекторе и у источника, соответственно, $\sigma_{\text{ISM}}$ — сечение поглощения приходящееся на один атом водорода и $N_{\text{H}} = \int_0^r n_{\text{H}} d\ell$ — количество водорода на луче зрения между источником и приемником. На больших расстояниях от Солнца $\sigma_{\text{ISM}}$ также будет зависеть от $\mathbf{n}$. Данная формула приведена не столько для того, чтобы напомнить закон поглощения, сколько для явного указания параметров от которых зависит данный процесс.

Удельное сечение поглощения в рассматриваемом нами диапазоне очень быстро убывает с ростом энергии $E$ и достаточно сильно зависит от химического состава МЗС, главным образом от содержания металлов $Z$ в межзвездном газе. Сечение умножается на лучевую плотность $N_H$, которая зависит как от расстояния $r$ до источника, так и от его положения на небе (которое обозначалось единичным вектором **n**). Рассмотрим каждую из величин подробнее.

В рассматриваемом спектральном диапазоне теоретические значения элементарных сечений фотоионизации различных атомов и молекул хорошо известны и вносят наименьшую ошибку в соотношение (9). Сечение поглощения $\sigma_{ISM}$ складывается из фотопоглощения на атомах различных элементов и молекулах, поглощения на пыли, поглощения и рассеяния на ионах и свободных электронах. В различных компонентах МЗС (холодной, теплой, горячей), оказывающихся на луче зрения, роль перечисленных процессов различны. Однако оценки показывают (Вилмс и др. 2000), что вкладами теплой и горячей (ионизованной) фаз МЗС можно пренебречь, поскольку большая часть межзвездной среды находится в слабоионизованном состоянии. Такое допущение нельзя делать для источников в остатках сверхновых или если на луче зрения оказывается мощная область ионизации.

В МЗС для энергий $E < 5$–$10$ кэВ основную роль в поглощении играют процессы фотоионизации и сечение поглощения $\sigma_{bf}$ быстро убывает с ростом $E$. Эта зависимость близка к $\sigma_{bf} \propto E^{-3}$, поскольку суммарный коэффициент поглощения состоит из совокупности фотоионизационных К–скачков для различных элементов, за каждым из которых сечение поглощения убывает приблизительно как $\propto E^{-3}$. Точнее, величина $\sigma_{bf} E^3_{keV}$ изменяется от $\sim 70 \cdot 10^{-24}$ см$^{-2}$ при $E = 0.1$ кэВ до $\sim 1000 \cdot 10^{-24}$ см$^{-2}$ при $E = 10$ кэВ (при солнечном химсоставе).

Суммарное сечение поглощения зависит от количества тяжелых элементов в МЗС. Наиболее простое предположение о химическом составе межзвездного газа является его совпадение с солнечным. Однако, измерения химического состава среды вне Солнечной системы указывают, что обилие тяжелых элементов там ниже, чем в Солнце и составляет $\sim 70$–$80\%$ от этого уровня для разных химических элементов (София и др. 1994, Саваж и Сембах 1996). Во столько же раз уменьшается сечение поглощения атомарного газа. Вероятно, основной причиной недостатка тяжелых элементов в МЗС по сравнению с Солнцем является их частичная конденсация в виде межзвездной пыли, которая в Солнце полностью испарена и перемешана с веществом.

Из межзвездных молекул следует принимать во внимание только молекулярный водород $H_2$. В среднем в МЗС $H_2$ составляет $20$–$25\%$ от H I (Грингел и др. 2000), однако это отношение обладает как сильной систематической зависимостью от расстояния от центра Галактики, так и существенными различиями на малых угловых масштабах. Указанное поведение вызвано тем, что основная доля $H_2$ заключена в молекулярных облаках, а большая часть таких облаков образует кольцо на расстоянии $\sim 5$ кпк от центра Галактики.

Поглощение, вызываемое межзвездной пылью, невелико, не более нескольких процентов от полного сечения $\sigma_{ISM}$. Его зависимость от $E$ примерно такая же, как и при фотоабсорбции. Исключение составляют ситуации, когда на луче зрения оказываются богатые пылью молекулярные облака или протозвездные объекты.

Последний нерассмотренный компонент — ионизованная фаза МЗС. Можно сразу сказать, что в МЗС для рентгеновского диапазона свободно–свободное поглощение неважно (см., например, Бочкарев 1992). А вот томсоновское рассеяние на электронах, которое практически не зависит от $E$, необходимо учитывать для $E > 10$ кэВ или для $N_H \gtrsim 10^{22}$ см$^{-2}$.

Типичные для Галактики значения лучевой плотности составляют $N_H = 10^{20}$–$10^{22}$ см$^{-2}$. Следует заметить, что величина $N_H$ может испытывать существенные локальные увеличения на малых угловых масштабах в отдельных точках небесной сферы — если поток излучения от рентгеновского источника пересекает компактное уплотнение МЗС, например, туманность, остаток сверхновой или молекулярное облако.

### 2.2.8 Популяционный синтез НЗ

В последнее время популяционный синтез стал популярным методом в различных областях астрофизики. В первую очередь он применяется для изучения эволюции двойных и одиночных звезд (см. Липунов и др. 1996, Фритце-ф.Альфенслебен 2000).

Обычно моделирование проводится методом Монте-Карло. Задаются начальные параметры объектов и законы их изменения. В идеале стремятся получить все эти данные в аналитической форме, однако часто используют затабулированные данные.

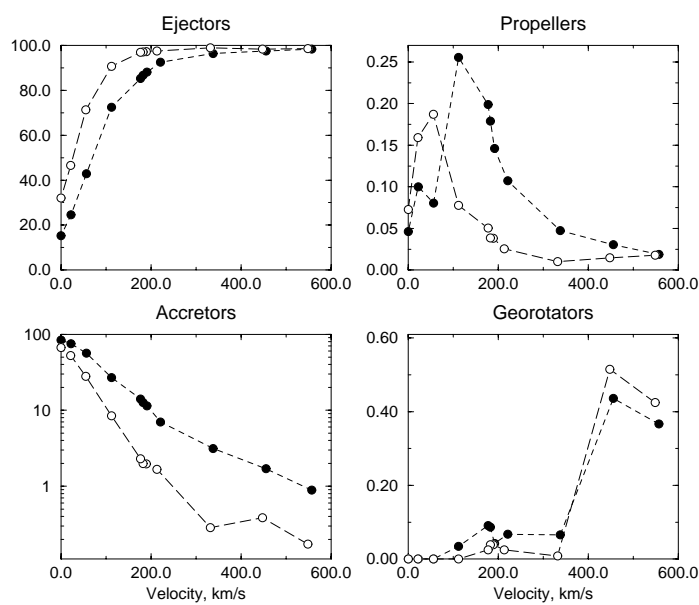

Рис. 10: Результаты популяционного синтеза одиночных НЗ (Попов и др. 2000а). Показано распределение НЗ по стадиям (Эжекторы, Пропеллеры, Аккреторы и Георотаторы) для постоянного магнитного поля в зависимости от средней скорости НЗ. Белые кружки — $\mu_{30} = 0.5$, черные — $\mu_{30} = 1$. $\mu$ — магнитный момент. По горизонтальной оси отложена средняя скорость НЗ (для максвелловского распределения).

Сравнение с наблюдениями может проводиться как на уровне параметров отдельных объектов, так и на уровне интегральных параметров, например, интегральный спектр галактики.

Как правило в популяционных моделях эволюционный трек отдельного члена популяции не имеет очень высокой точности. Однако, большая совокупность объектов позволяет судить о таких важных параметрах, как распределение начальных параметров, а также делать предсказания, касающиеся популяции в целом (см., например, работы, посвященные пространственному распределению одиночных НЗ в Галактике в связи с гипотезой о галактическом происхождении гамма-всплесков: Пачинский 1990, Прохоров, Постнов 1993, 1994).

Метод популяционного синтеза дает существенные преимущества по сравнению с исследованием свойств индивидуальных объектов (см. например Липунов и др. 1996). Различные популяции изолированных НЗ не раз изучались с помощью данного метода (см. Бхатачарья и др. 1992 , Мадау и Блаез 1994, Мэннинг и др. 1996). Детальное изучение НЗ в применении к слабым рентгеновским источникам в диске Галактики и в шаровых скоплениях было недавно проведено в (Попов и др. 2000б, Попов, Прохоров 2002). Исследовался вклад одиночных аккрецирующих НЗ в рентгеновский фон (Дзане и др. 1995), а также НЗ в центре Галактики (Дзане и др. 1996).

В работе Попов и др. 2000а было получено распределение НЗ по стадиям (Эжектор, Пропеллер, Аккретор, Георотатор). Расчеты показали, что большая часть НЗ находится на стадии Эжекции. Это связано с высокими (в среднем) пространственными скоростями этих объектов (Лайн и Лоример 1994, Кордес и Чернофф 1997, 1998, Хансен и Финней 1997,

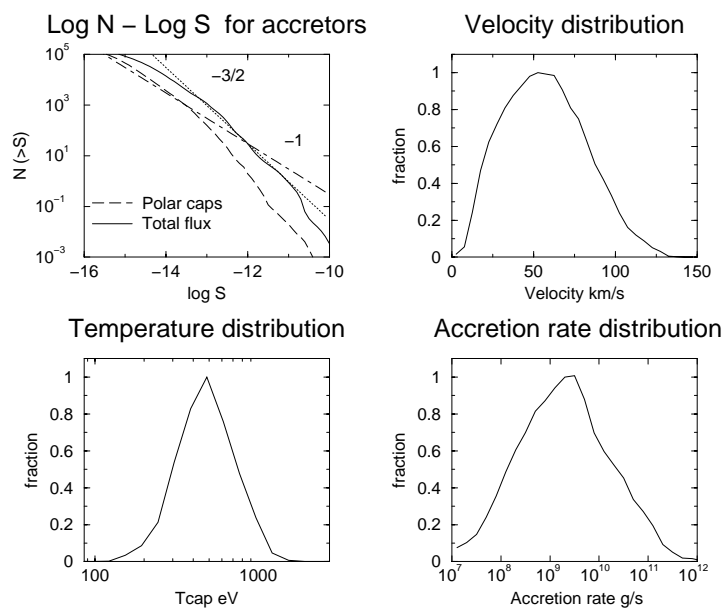

Рис. 11: Результаты популяционного синтеза одиночных аккрецирующих НЗ (Попов и др. 2000б). На верхнем левом рисунке приведены кривые Log N − Log S для полного потока (рассчитанного по формуле Бонди и 100-процентной эффективности аккреции) и для аккреции на полярные шапки в диапазоне 0.5-2 кэВ. Магнитное поле предполагалось постоянным с распределением, соответствующим наблюдающемуся у радиопульсаров. Распределение по скоростям было максвелловским со средним значением 300 км/с. На трех других рисунках показано распределение аккрецирующих НЗ по скоростям, температуре излучения и темпу аккреции.

Хартман 1997, Рамачандран 1999, Кордес 1998, см. также Тревес и др. 1998). Аккреторы составляют порядка нескольких процентов от всей популяции в случае постоянного магнитного поля.

Эволюция двойных и одиночных радиопульсаров неоднократно исследовалась методом популяционного синтеза. Здесь мы не будем рассматривать миллисекундные радиопульсары (см. Поссенти и др. 1999), обратимся к эволюции одиночных пульсаров, следуя серии работ (Вербунт и др. 1999, Бхатачарья и др. 1992, Хартман и др. 1997).

Одним из ключевых вопросов в эволюции НЗ является проблема распада магнитного поля. Если в двойных системах распад может быть ускорен аккрецией вещества (см. Гепперт и др. 1996), то в одиночной НЗ распад обусловлен только внутренними причинами. Поэтому именно исследование изолированных объектов представляет особый интерес (Попов, Прохоров 2000). Поскольку до недавнего времени только радиопульсары удовлетворяли этому требованию, не удивительно, что авторы работ (Вербунт и др. 1999, Бхатачарья и др. 1992, Хартман и др. 1997) обратились именно к ним.

Специфическая трудность заключена в сравнении результатов с наблюдениями. Дело в том, что пульсарные данные подвержены множеству селекционных эффектов. Для сравнения расчетов с данными наблюдений необходимо ввести критерий детектируемости для моделируемых НЗ. Кроме этого, данные по расстояниям до пульсаров не обладают высокой степенью надежности (например, в 1994 году они были существенно пересмотрены (Лайн, Лоример 1994)).

Кроме изменения периода вращения и величины магнитного поля в расчетах необходимо учитывать движение НЗ в Галактике, т.к. за время жизни пульсар может существенно удалиться от места своего рождения.

Наилучшего согласия с данными наблюдений авторы достигли используя следующий набор параметров: $\tau \sim 10^8$ лет, $\log B_0 = 12.34$, $\sigma_B = 0.34$. Здесь $B = B_i \exp(-t/\tau)$, а распределение начальных полей соответствует распределению $B = 1/(\sqrt{(2\pi)}\sigma_B) \cdot exp(-1/2\,((\log B_i - \log B_0)/\sigma_B))$.

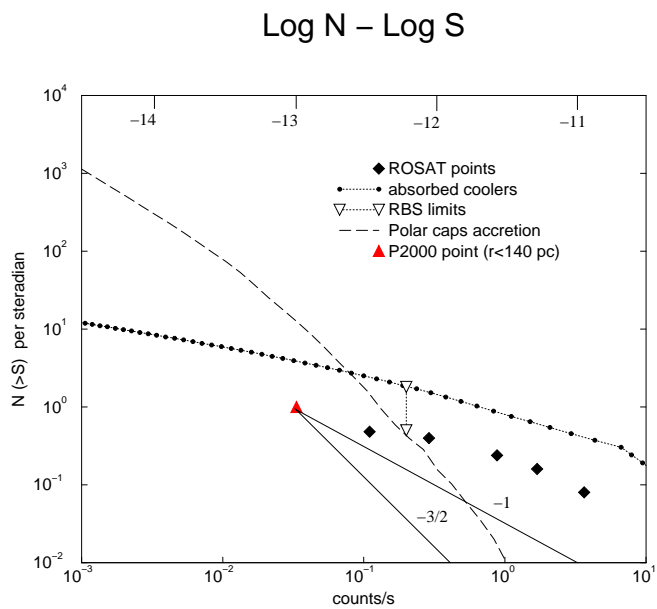

Рис. 12: Диаграмма Log N — Log S (из работы Попов 2001, с исправлениями). На горизонтальной оси отложены отсчеты в секунду для спутника РОСАТ. Ромбами показаны объекты "великолепной семерки". Треугольники — предел из РОСАТовского каталога ярких источников. Пунктирная кривая — результаты расчетов для аккрецирующих НЗ с постоянным магнитным полем, аккреция идет на полярные шапки. Точки – кривая для молодых охлаждающихся НЗ в предположении их высокой пространственной плотности в окрестностях Солнца. Черным треугольником показана точка, соответствующая расчетам Попова и др. (2000а), для наглядности от нее проведены прямые с наклоном -1 и -3/2. Видно, что аккрецирующие НЗ с постоянным магнитным полем не могут объяснить наблюдения. Требуется повышенная плотность молодых НЗ (это может объясняться Поясом Гулда, см. Попов и др. 2002) для объяснения наблюдаемой кривой Log N — Log S.

Последняя (по времени) попытка популяционного синтеза радиопульсаров с учетом затухания магнитного поля приведена в работе (Регимбо и де Фрейтас Пачеко 2001). Авторы также пришли к выводу о том, что наличие затухания на временах, превышающих время жизни пульсара, способствует улучшению статистики. Однако, в работе показано, что лучших результатов можно достичь и без затухания, если в процессе эволюции НЗ увеличивается угол между магнитной осью и осью вращения (см. Бескин и др. 1993, где приводятся теоретические аргументы в пользу такой эволюции). Кроме этого, авторы делают вывод о том, что наличие магнитаров может быть объяснено без бимодальности функции распределения НЗ по магнитным полям, и оценивают темп рождения магнитаров как 1/750 лет.

Кроме исследования популяции радиопульсаров в целом различные авторы рассматривали важный вопрос об ассоциации этих объектов с остатками сверхновых. Впервые к вопросу о малом числе пар (ассоциаций) обратились в конце 80-х (Браун и др. 1989, Нараян, Шаудт 1988). Нараян и Шаудт предположили, что пульсары не наблюдаются в некоторых остатках из-за того, что магнитное поле слишком слабо или период вращения НЗ слишком велик, т.е. пульсар находится или очень близко к линии смерти или же уже за ней (о линии смерти см. Чен и Рудерман 1993, Аронс 2000). Более современные расчеты были проведены в работах Генслера и Джонстона (1995а,б,в). Основной вывод работы Генслера и Джонстона (1995в) заключается в том, что даже если каждая сверхновая порождает радиопульсар, то число ассоциаций будет меньше наблюдаемого. Т.е. среди имеющихся пар пульсар+остаток часть является лишь результатом проекции.

В настоящее время в России не проводятся исследования по популяционному синтезу радиопульсаров.

Методом популяционного синтеза было исследовано распределение компактных объектов по массам (Бельчинский и др. 2002). Кроме тривиального результата о наличии попу-ляции черных дыр в двойных системах с более высокими массами по сравнению с массами одиночных черных дыр, авторы нашли, что количество странных звезд (см. также ниже) оказывается порядка количества черных дыр. При этом большая часть странных звезд является одиночными объектами.

В связи с популяционным синтезом НЗ особенно интересны попытки построения "единых теорий", описывающих с единых позиций НЗ разных типов. Некоторые такие попытки рассмотрены в разделе, посвященным аккреции, другие — в разделе, посвященным магнитарам. Ниже мы кратко суммируем эти работы.

Как было указано выше существует две основные ветви: "магнитарная" и "аккреционная".

В первом случае предполагается, что различия в проявлениях НЗ определяются разной величиной магнитного поля (и/или различной конфигурацией поля). НЗ с большими полями могут в принципе не проявлять радиопульсарной активности (Баринг и Хардинг 1995, см., однако, Усов и Мелроуз 1996). Возможно, также, что из-за быстрого замедления стадия радиопульсара для этих звезд оказывается очень короткой. "Магнитарная" гипотеза обязательно привлекает затухание магнитного поля для объяснения кластеризации периодов МПГ и АРП (Колпи и др. 2000). Различие между источниками может определяться разницей в возрасте, и, соответственно, различием в величине поля.

Во втором случае для единого объяснения природы различных одиночных НЗ требуется наличие остаточных аккреционных дисков. В таком случае различные источники объясняются или как последовательные стадии (эволюция диска и темпа вращения НЗ), или как НЗ с разными параметрами дисков.

Хорошую дискуссию по этим проблемам можно найти в работах Томпсона (2001), Альпара (2001), Дункана (2001).

## 2.3 Другие близкие области исследований

### 2.3.1 Механизмы взрывов сверхновых и возвратная аккреция (fall-back)

Поскольку НЗ является одним из результатов взрыва сверхновой, то, разумеется, свойства молодых НЗ несут информацию о породивших их сверхновых (см. Имшенник 2000, 1998; Янка и др. 2001).

Примером такой связи является проблема начальных скоростей НЗ (kick) и начальных периодов. В настоящий момент распределение начальных скоростей НЗ неизвестно. Наблюдения радиопульсаров (Лайн, Лоример 1994, Лоример и др. 1997) указывают на высокие средние скорости. Распределение по начальным периодам также неизвестно. Пока достоверно известны начальные периоды для пульсара в Крабе и, возможно, еще для двух пульсаров (см. Каспи, Хельфанд 2002). Как предполагают Спруит и Финней (1998) и Постнов и Прохоров (1998), распределение по начальным периодам может быть связано с процессом взрыва сверхновой и, соответственно, с распределением по скоростям. Однако, в работах этих авторов сделаны выводы *противоречащие* друг другу.

В работах по взрывам коллапсирующих сверхновых (типа II и Ib/c) можно встретить три качественно различных механизма преобразования выделяющейся гравитационной энергии связи коллапсирующего ядра предсверхновой в кинетическую энергию сбрасываемой оболочки.

Первый механизм объединяет классическую модель "отскока" ("bounce") падающих внешних слоев ядра предсверхновой от сформировавшегося и ставшего жестким сверхплотного остатка сверхновой (горячей прото-НЗ) с нейтринными механизмами, в которых образовавшаяся в результате отскока ударная волна в дальнейшем подпитывается нейтринным излучением горячего ядра. Это самая первая и долгое время считавшаяся основной модель взрыва сверхновой. Хотя ранее в рамках этой модели несколько раз удавалось объяснить вспышку сверхновой, последующие более точные исследования отвергали эти найденные возможности (см. Меццакаппа и др. 1998а и ссылки там). На сегодня данный механизм не объясняет сброс оболочки сверхновой ни в сферически симметричном, ни в осесимметричном (с вращением) случаях (Янка и др.2001). Есть надежда, что ситуацию могла бы исправить крупномасштабная нейтринная конвекция (Херант и др. 1994, Меццакаппа и др. 1998б). В настоящее время в данном направлении ведутся интенсивные исследования (см. например Кифонидис и др. 1999 и ссылки там).

Другой механизм (Имшенник 1992) связан с делением быстровращающегося коллапсирующего ядра звезды на 2 части, по крайней мере одна из которых должна быть нейтронной звездой. Затем части двойного ядра сближаются из-за гравитационного излучения, пока меньшая по массе (и большая по размеру) компонента не заполнит свою полость Роша. Сближение двойного ядра может длиться от нескольких минут до нескольких часов. После этого начинается перетекание вещества до тех пор, пока масса меньшей компоненты не достигнет нижнего предела масс НЗ (около $0.09 M_\odot$), при котором происходит взрывная денейтронизация маломассивной нейтронной звезды (Блинников и др. 1984). Такое дополнительное выделение энергии, происходящее достаточно далеко от центра коллапсирующей звезды, может эффективно сбросить ее оболочку. Этот механизм может действовать только у наиболее быстро вращающихся предсверхновых. Проблема данного сценария заключена в том, что пока еще никому не удалось воспроизвести деление ядра предсверхновой при коллапсе.

Последний из рассматриваемых нами механизмов взрывов сверхновых — магниторотационный — был предложен Г.С.Бисноватым-Коганом в 1970 г. Идея этого механизма очень проста — сброс оболочки производится магнитным полем быстро вращающейся НЗ. При этом оболочка ускоряется за счет торможения вращения нейтронной звезды. Поскольку на самом деле эта простая идея объединяет в себе генерацию и усиление магнитных полей и сложную трехмерную гидродинамику с сильным влиянием переноса излучения, то реалистичные расчеты данного сценария крайне затруднены. Результаты двумерных расчетов (Арделян и др. 1998, 2000) показывают, что магниторотационный механизм может передать несколько процентов вращательной энергии компактного остатка в кинетическую энергию оболочки. Как показывают упомянутые расчеты, магниторотационный взрыв (стадия на которой происходит существенное ускорение и сброс оболочки) длится 0.01–0.1 с. Однако ему предшествует фаза "накрутки", на которой тороидальное магнитное поле линейно усиливается до критической величины ($\sim 10^{16}$–$10^{17}$ Гс) при которой происходит сброс оболочки. Длительность этой стадии зависит от величины начального магнитного поля НЗ и от скорости ее вращения и может меняться от долей секунды до минут (и даже часов). Для данного механизма требуется достаточно быстрое вращение НЗ (период порядка нескольких миллисекунд), однако не столь быстрое, как в механизме с делением ядра.

При рассмотрении связи НЗ и вспышек сверхновых особенно интересной оказывается проблема fall-back, падения на НЗ вещества остатка сверхновой (Хук, Шевалье 1991). Существует ряд моделей (см. Альпар 2001), объясняющих природу АРП, МПГ, слабых рентгеновских источников в диске Галактики и компактных рентгеновских источников в остатках сверхновых выпадением вещества, выброшенного при взрыве, на поверхность НЗ (см. пункт, посвященный аккреции).

Выпадение вещества на образовавшийся в результате взрыва сверхновой компактный объект рассматривалось уже в начале 70-х годов (Колгейт 1971, Зельдович и др. 1972). В последние годы были получены новые важные результаты (см. Шевалье 1989, Зампьери и др. 1998). Получены оценки времени, после которого аккреционная светимость образовавшейся черной дыры превосходит другие источники излучения. Мониторинг известных сверхновых позволит в течение нескольких лет проверить эту модель.

### 2.3.2 Одиночные черные дыры

Аккрецирующие одиночные НЗ могут быть довольно близки по многим параметрам к одиночным аккрецирующим черным дырам (о физике черных дыр см. книгу Новикова, Фролова 1986). Различные аспекты, связанные с аккрецией на одиночные черные дыры звездных масс изучались уже 30 лет назад (см. Шварцман 1971).

Недавно аккреция из МЗС на одиночные черные дыры была рассмотрена Фуджитой и др. (1998) и Аголом и Камионковским (2001). Авторы рассмотрели как обычные черные дыры звездных масс, так и гипотетические объекты, которые могут вносить существенный вклад в темную массу гало нашей Галактики, обнаруженные по эффекту микролинзирования (см. также Агол и др. 2002 об обнаружении кандидатов в черные дыры методом микролинзирования). Используя модель адвекционно-доминированной аккреции, было показано, что черные дыры звездных масс могут в недалеком будущем наблюдаться в рентгеновском, ИК или оптическом диапазонах. Обнаружение менее массивных объектов гало за счет аккреции маловероятно, однако отмечается вероятность обнаружения гравитационных волн от слияния таких черных дыр, если они образуют достаточно тесные двойные системы.

Балберг и Шапиро (2001) рассчитали темп образования черных дыр после вспышек сверхновых. Используя аналитическую модель изменения аккреционной светимости (за счет fall-back, см. выше) они оценили, что при современных методах наблюдений можно до нескольких "проявлений" черных дыр в год. Т.е., наблюдая кривые блеска сверхновых, можно увидеть как начинает доминировать вклад аккреции в полную светимость сверхновой. Если эти оптимистические оценки оправдаются, то в скором будущем мы сможем непосредственно определить темп рождения черных дыр.

Вероятно существует возможность обнаружить одиночные черные дыры в непосредственной окрестности Солнца, на расстоянии менее 1 кпк. Такая возможность связана с существование массивных "убегающих" звезд. Эти объекты образуются в результате распада двойных систем. Высокая масса ($> 30\, M_\odot$) убегающих звезд говорит о том, что образовавшимся в результате взрыва второго компонента компактным объектом является черная дыра (см., однако, статьи Эргмы и ван ден Хевела 1998а,б о массе прародителей НЗ и черных дыр). Таким образом оказывается возможным вычисление приближенных положений близких черных дыр (Попов и др. 2002, Прохоров, Попов 2002). Другая возможность поиска одиночных черных дыр на основе обработки массовой многоцветной фотометрии (обзора SDSS) подробно рассмотрена в работе Чисхолм и др. (2002).

### 2.3.3 Странные звезды

В ядре НЗ плотность может существенно (в несколько раз) превосходить ядерную. Это создает условия для существования свободных кварков (деконфайнмент). Такая возможность была осознана в начале 70-х годов. Первой работой, посвященной кварковому веществу была статья Бодмер (1971). Первыми работами, посвященными непосредственно кварковым звездам, были статьи Фечнера и Джосса (1978) и Виттена (1984) (детальнее см. обзор Бомбачи 2001).

Странные звезды имеют уравнение состояния, отличное от нормальных НЗ. Это проявляется, в частности, в меньших радиусах при той же массе компактного объекта. Таким образом измерения массы и радиуса объекта могут позволит определить его природу. С этой точки зрения чрезвычайно актуальны наблюдения одного из семи объектов "великолепной семерки" — RX J1856 (см. Понс и др. 2001). Пачинским (2001) была предложена идея определения массы этого объекта с помощью микролинзирования, которая может быть осуществлена в ближайшие несколько лет (однако, если верна правильная оценка расстояния до этого объекта, сделанная Капланом и др. (2001б), а не оценка Волтера (2001), то идея Пачинского об определении массы НЗ *не сможет быть осуществлена* в ближайшие годы). Более подробное исследования микролинзирования на НЗ было недавно проведено Шварцем и Сейделом (2002). В будущем возможны одновременные измерения массы и радиуса НЗ при наблюдении гравитационных волн (Валлиснери 2002).

В настоящее время существует несколько кандидатов в странные звезды в тесных двойных системах: 4U 1820-30, SAX J1808.4-3658, 4U 1728-34, Her X-1, GRO J1744-28 (см. Бомбачи 2002). Некоторые авторы связывают активность МПГ со странными звездами (см., например, Дар, Де Рухула 2000, Усов 2001). Однако, основная доля исследований, посвященных странным звездам, представлена теоретическими работами. В России теоретические исследования в этой области активно ведутся Д.Н. Воскресенским (МИФИ) и соавторами (см. Блашке и др. 2001 и ссылки там, а также обзор Воскресенского в материалах конференции "Physics of neutron star interiors" 2001).

# 3 Заключение

В заключение еще раз повторим, что, благодаря прогрессу в наблюдательной астрономии в последние 10 лет, складывается картина, в которой значительная роль принадлежит *радиотихим* НЗ.

По всей видимости значительная доля НЗ не проходит в молодости стадию радиопульсара, или же эта стадия оказывается очень короткой. Около 10% НЗ могут являться магнитарами. Часть НЗ может сохранять существенные остаточные аккреционные диски за счет выпадения вещества из оболочки (fall-back). Этот же процесс может существенно влиять на начальные параметры НЗ.

Картина эволюции НЗ уже не выглядит такой ясной, как 15-20 лет назад, когда пульсар в Крабе считался "идеальной молодой НЗ". Для дальнейшего прогресса необходимы совместные усилия наблюдателей и теоретиков.

## Список литературы